\documentclass[aps, prx,twocolumn,showpacs,preprintnumbers,amsmath,amssymb,a4paper,longbibliography]{revtex4-1}

\usepackage{mathrsfs}%hanamoji
\usepackage{graphicx}% Include figure files
\usepackage{dcolumn}% equation table columns on decimal point
\usepackage{bm}% bold math
\usepackage{here}
\begin{document}
\title{Linear irreversible thermodynamics and Onsager reciprocity for information-driven engines}
\author{Shumpei Yamamoto$^1$, Sosuke Ito$^2$, Naoto Shiraishi$^1$, and Takahiro Sagawa$^3$}
\affiliation{$^1$Department of Basic Science, The University of Tokyo,3-8-1 Komaba, Meguro-ku, Tokyo 153-8902, Japan\\
 $^2$Department of Physics, Tokyo Institute of Technology, Oh-okayama 2-12-1, Meguro-ku, Tokyo 152-8551, Japan\\
$^3$ Department of Applied Physics, The University of Tokyo, 7-3-1 Hongo, Bunkyo-ku, Tokyo 113-8656, Japan
}
\date{\today}
\begin{abstract}
In the recent progress in nonequilibrium thermodynamics, information has been recognized as a kind of thermodynamic resource that can drive thermodynamic current without any direct energy injection. In this paper, we establish the framework of linear irreversible thermodynamics for a broad class of autonomous information processing. In particular, we prove that the Onsager reciprocity holds true with information: The linear response matrix is well-defined, and is shown symmetric with both of the information affinity and the conventional thermodynamic affinity.
As an application, we derive a universal bound for the efficiency at maximum power for information-driven engines in the linear regime. Our result reveals the fundamental role of information flow in linear irreversible thermodynamics.
\end{abstract}

\maketitle

\section{INTRODUCTION}

Linear irreversible thermodynamics has played important roles in physics, chemistry, and quantitative biology. In particular, the Onsager reciprocity~\cite{Onsager1,Onsager2}, characterizing the symmetry of linear transport coefficients, gives deep insight into a variety of phenomena, from the thermoelectric effect~\cite{Callen, GoupilSnyder, BrandnerSeifert} to membrane transport processes in biological systems~\cite{KedemKatchalsky,MasonLonsdale,Baranowski,Richard,Gerber}.
The Onsager reciprocity can be placed in the framework of stochastic thermodynamics~\cite{Schnakenberg,Sekimoto,Seifert}, which is an extension of thermodynamics to small fluctuating systems. 
More precisely, the Onsager reciprocity and its nonlinear generalization can be derived from the fluctuation theorem~\cite{Andrieux,Andriuex2,Andrieux3,SaitoUtsumi}.
The Onsager reciprocity also plays a central role in the finite-time thermodynamics: The efficiency of thermodynamic engines at maximum power (i.e., the work per unit time)~\cite{CurzonAhlborn} has been studied on the basis of the Onsager reciprocity~\cite{Broeck,BenentiCasati}.
Furthermore, novel properties of the linear response matrix beyond the Onsager reciprocity have been found for periodically driven systems~\cite{ProesmansBroeck,BrandnerSeifert2}.

Stochastic thermodynamics has been further extended to information processing such as measurement and feedback control, which accompanies the refined second law of thermodynamics by taking information into account~\cite{review,TouchetteLloyd,SagawaUeda08,SagawaUeda09,AllahverdyanMahler,SagawaUeda2,FujitaniSuzuki,HorowitzVaikuntanathan,SagawaUeda4}.
This research direction 
sheds new light on the problem of ``Maxwell's demon,'' and the demon has experimentally been investigated
\cite{Toyabe,Berut,Koski}.
Thermodynamics of information has also been applied to biological signal transduction and adaptation processes~\cite{MehtaSchwab2,ItoSagawa2,HartichSeifert,SartoriHorowitz,tenWolde,Lan,Mehta,LangMehta}.
Furthermore, thermodynamics of autonomous information processing has attracted much attention, where an autonomous Maxwell's
 demon reduces the entropy of an engine by continuous measurement and feedback control without any direct energy exchange~\cite{MandalJarzynski,StrasbergEsposito,HorowitzParrondo2,DeffnerJarzynski,ItoSagawa,HorowitzEsposito,Hartich,Munakata,HorowitzSandberg,ShiraishiSagawa,ShiraishiItoKawaguchiSagawa,ShiraishiMatsumotoSagawa,Koski2, HartichSeifertSensor,BaratoSeifert1,BaratoSeifert2,BaratoSeifert3}.
Such a thermodynamic system is called an autonomous information-driven engine.
It has been shown that continuous information flow can be treated on an equal footing with thermodynamic currents, and the concept of information affinity has been introduced~\cite{HorowitzEsposito}.
However, general linear irreversible thermodynamics has not been addressed in these previous researches.
In particular, it has been a fundamental open question whether the Onsager reciprocity is still valid in the presence of the
 information flow and the information affinity.

In this paper, we establish linear irreversible thermodynamics for a broad class of autonomous heat engines in the presence of continuous information flow. 
We show that the Onsager reciprocity is indeed valid with the information affinity, which implies a nontrivial symmetry between thermodynamic and informational currents and affinities.
As a special application, we derive the information-thermodynamic efficiency at maximum power of information-driven engines
~\cite{BauerSeifert,SandbergMitter,JarilloCao,BaratoSeifert2}, which, in the linear regime, is universally bounded by the half of the maximum efficiency.

Our linear irreversible thermodynamics is applicable if only a driven engine is close to equilibrium, even when the entire system including Maxwell's demon is far from equilibrium.
This makes a sharp contrast to conventional linear irreversible thermodynamics, where the entire system must be close to equilibrium.
This feature of our framework is based on  a fundamental lemma that  is proved in this paper, which states that if all affinities including the information affinity are zero, all the conjugate currents are zero. 
This lemma ensures that linear irreversible thermodynamics is a consistent framework even in the presence of the information affinity. 
%We note that our formulation is based on the Schnakenberg network theory~\cite{Schnakenberg}.
%Our results would be useful to analyze autonomous information processing in biological systems.**
%, such as sensory adaptation processes.

%A crucial characteristic of our linear irreversible thermodynamics lies in the fact that it is applicable if only a driven engine is close to equilibrium, even when the entire system is far from equilibrium.
%In particular, our theory  is applicable to information engines driven by Maxwell's demon, even when the demon itself is far from equilibrium.

We note that there is previous work about the Onsager reciprocity for heat engines driven by ``information reservoirs''~\cite{BaratoSeifert2, BaratoSeifert3}.
In that work, however, the role of information flow and information affinity was not taken into account.
Furthermore, the entire system must be close to equilibrium in their setup, as is the case for conventional linear thermodynamics without information.
Our work establishes linear irreversible thermodynamics including continuous information flow.

Our result would be applicable to analyze the role of information in biological systems.
In fact, a variety of phenomena in biological systems are found in the linear nonequilibrium regime~\cite{Demirel}.
Membrane transport is one such phenomena~\cite{KedemKatchalsky,MasonLonsdale,Baranowski}, where the Onsager reciprocity has experimentally been verified~\cite{Richard}.
Recently, the Onsager coefficient has been determined from experimental data of living yeast~\cite{Gerber}.
We also note that biochemical information processing has been studied with stochastic models similar to our setup~\cite{SartoriHorowitz,Mehta,BaratoSeifert1}.
We emphasize that our theory is applicable if only some biochemical reactions are close to equilibrium, even when an entire biological system is far from equilibrium.

Throughout this paper, we assume that a system obeys a continuous-time Markov jump process with finite states.
We also assume that the system has the time-reversal symmetry, where any variable does not change its sign by the time-reversal transformation.
In particular, we assume that there is no magnetic field.

This paper is organized as follows.
In Sec.~\ref{paradigmatic}, we discuss the essentials of our results with a simple model of autonomous information-driven engines, which we call the four-state model.
We introduce the four-state model in Sec.~\ref{setup}, and review previous results on thermodynamics of information in Sec.~\ref{second law}. We then state a specific form of our main results for the four-state model in Sec.~\ref{onsager coefficient}. We apply our results to derive the efficiency at maximum power in Sec.~\ref{ssmaxpower}.
In Sec.~\ref{generalbipartite}, we formulate the general setup and discuss our main results.
We formulate the setup in Sec.~\ref{setupG}, and review thermodynamics of information in Sec.~\ref{partial affinity}.
We present the Onsager reciprocity in the general form in Sec.~\ref{ORRinG}, and prove it in Sec.~\ref{proof}.
In Sec.~\ref{conclusion}, we conclude this paper with some remarks.
In Appendix A, we prove the aforementioned technical lemma.
In Appendix B, we show the explicit  steady distribution of the four-state model.
 
\begin {figure}[]
\centering
\includegraphics[width=8cm]{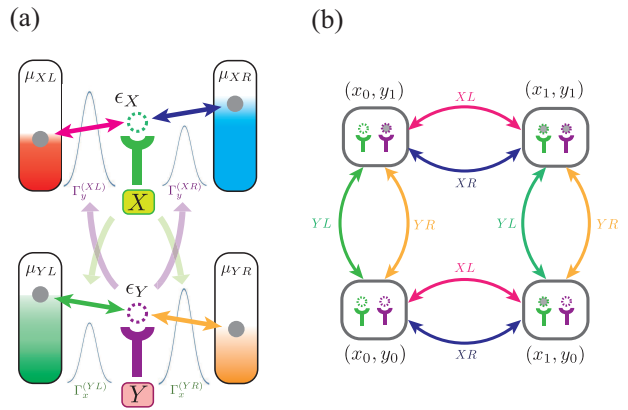}
\caption{(Color online)(a) Schematic of the four-state model. A single-particle site of $X$ ($Y$) exchanges particles with two particle reservoirs $XL$ and $XR$ ($YL$ and $YR$) with chemical potentials $\mu_{XL} $ and $\mu_{XR}$ ($\mu_{YL}$ and $\mu_{YR}$). The height of the barriers between $X$ ($Y$) and the two reservoirs are characterized by $\Gamma_y^{(XL)}$ and $\Gamma_y^{(XR)}$ ($\Gamma_x^{(YL)}$ and $\Gamma_x^{(YR)}$), which depend on the state of $Y$ ($X$). (b) State space of the four-state model. Vertices (i.e., nodes) represent the four states, and bidirected edges represent the forward and backward transitions between two states induced by particle reservoirs with nonzero transition rates.}
\label{4state}
\end{figure}

\section{\label{paradigmatic}A PARADIGMATIC MODEL}
Before going to the general argument, we illustrate the essentials of our results by considering a minimal model of information-driven autonomous engines, which we call the four-state model.
This model consists of two subsystems with respectively two states, and thus the whole system has four possible states.
The two subsystems autonomously interact with each other, where they do not directly exchange energy but exchange information.
Therefore, one of the two subsystems is regarded as a Maxwell's demon that continuously measures the other system and performs feedback control.
Our model is equivalent, or almost equivalent, to well-studied autonomous information engines discussed in Refs.~\cite{Sekimoto,SekimotoSE, Lan, Mehta, StrasbergEsposito,ShiraishiSagawa,BaratoSeifert1, HorowitzEsposito,Schaller}, which are closely related to transport in biological systems.

\subsection{\label{setup}Setup}

The four-state model is a bipartite system that consists of two subsystems $X$ and $Y$. We refer to the composite system as $Z$.
$X$ ($Y$) is attached to two particle reservoirs $XL$ and $XR$ ($YL$ and $YR$) at inverse temperature $\beta$.
Each system has a single site where a particle comes in from or goes out to one of the attached particle reservoirs. 
The chemical potential differences between $X$ ($Y$) and the particle reservoirs are given by $\mu_{XL}$ and $\mu_{XR}$ ($\mu_{YL}$ and $\mu_{YR}$) (see also Fig.~\ref{4state}). 
The energy of $X$ $(Y$) is $\epsilon_X$ ($\epsilon_Y$) when it is filled with a particle, and zero when it is empty.

The state of each system is represented by the number of particles in the site (i.e., filled or empty).
The entire states of composite system $Z$ are labeled as $z=(x_0, y_0), (x_1, y_0), (x_0, y_1), (x_1, y_1)$, where $x_0$ and $y_0$ ($x_1$ and $y_1$) represent that the site of $X$ and $Y$ is empty (filled), respectively.

Let $p(z, t)$ be the probability of state $z$ of the composite system at time $t$. The time evolution of $p(z,t)$ is described by the master equation:
\begin{align}
\frac{d}{dt} p(z,t) = \sum_{z', \nu} [W_{z z'}^{(\nu)} p(z',t) - W_{z'z }^{(\nu)} p(z,t) ],
\label{mastereq}
\end{align}
where $W_{z z'}^{(\nu)}$ is the time-independent transition rate from $z'$ to $z$ induced by reservoir $\nu \ (=XL, XR, YL, YR)$.
We assume that subsystems do not change their own states simultaneously in a single transition, which is equivalent to the bipartite condition of the transition rates:
\begin{align}
W^{(\nu)}_{z z'} =
\begin{cases}
W^{(\nu)}_{x x' |y}\ &(x\neq x', y=y', \nu=XL,\ XR)\\
W^{(\nu)}_{y y' |x}\ &(x=x', y\neq y', \nu=YL,\ YR)\\
0 &( {\rm otherwise}).
\end{cases}
\label{bipartite1}
\end{align}

 We assume that the transition rates satisfy the local detailed balance conditions:
\begin{align}
\frac{W^{(\nu)}_{x_1 x_0 |y_i}}{W^{(\nu)}_{x_0 x_1 |y_i}} = \exp (- \beta (\epsilon_X - \mu_{\nu}) ), \nonumber\\
\frac{W^{(\nu)}_{y_1 y_0 |x_i}}{W^{(\nu)}_{y_0 y_1 |x_i}} = \exp ( - \beta (\epsilon_Y - \mu_{\nu}) ). 
\label{detailed_balance}
\end{align} 
Our theory is applicable independent of the details of the transition rates, as long as they satisfy Eq.~(\ref{detailed_balance}).
As a specific example, the transition rates can be of the form
\begin{align}
W^{(\nu)}_{x_1 x_0 |y_i} &= \Gamma^{(\nu)}_{y_i} f^{(\nu)}_X,\nonumber\\
W^{(\nu)}_{x_0 x_1 |y_i} &= \Gamma^{(\nu)}_{y_i} (1 -f^{(\nu)}_X ),\nonumber\\
W^{(\nu)}_{y_1 y_0 |x_i} &= \Gamma^{(\nu)}_{x_i} f^{(\nu)}_Y,\nonumber\\
W^{(\nu)}_{y_0 y_1 |x_i} &= \Gamma^{(\nu)}_{x_i} (1 -f^{(\nu)}_Y ),
\label{bipartite1_1}
\end{align}
 where $f^{(\nu)}_X := [1+\exp ( \beta (\epsilon_X - \mu_{\nu}) ) ]^{-1}$ and $f^{(\nu)}_Y := [1+\exp (\beta (\epsilon_Y - \mu_{\nu}) ) ]^{-1}$ are the Fermi distribution functions, and $\Gamma^{(\nu)}_{y_i}$ ($\Gamma^{(\nu)}_{x_i}$) is a positive constant that characterizes the height of the potential barrier between $X$ ($Y$) and reservoir $\nu$. 
 The potential barrier of $X$, characterized by $\Gamma^{(\nu)}_{y_i}$, depends on the state of $Y$, and vice versa.

As mentioned above, the transition rates between a site and the attached reservoirs are determined by the other site.
On the other hand, the two subsystems do not directly exchange energy, because the energy of each site is independent of the state of the other.
These assumptions make it reasonable to call our system as an information-driven engine.
Intuitively, each subsystem continuously measures the other system, and performs feedback control by changing the energy barrier between the other system and the attached reservoirs. 
In the special case that $\Gamma^{(XL)}_{y_0}=1,\ \Gamma^{(XR)}_{y_0}=0,\ \Gamma^{(XL)}_{y_1}=0$, and $\Gamma^{(XR)}_{y_1}=1$, $X$ exchanges a particle only with $XL$ ($XR$) when $Y$ is empty (filled), respectively.
This is a typical model of an autonomous Maxwell's demon, where the demon inserts a wall between the engine and reservoirs, depending on the number of particles in the engine~\cite{ShiraishiSagawa}.

\subsection{\label{second law}Second law and information flow}
In this subsection, we briefly review previous results on the generalized second law including information flow, which has been discussed in Ref.~\cite{HorowitzEsposito}.
Let $p_{\rm ss}(z)$ be the steady state distribution
of the four-state model. The thermodynamic affinity of $X$ is characterized by the chemical potential difference between $XL$ and $XR$ as
\begin{equation}
F_X := \beta (\mu_{XL}-\mu_{XR}).
\label{thermo_affinity}
\end{equation}
 The net particle current from $XL$ to $XR$, which is conjugate with $F_X$, is given by
\begin{align}
J_X:=&W^{(XR)}_{x_0 x_1|y_0}p_{\rm ss}(x_1,y_0)-W^{(XR)}_{x_1 x_0|y_0}p_{\rm ss}(x_0,y_0)\nonumber\\
&+W^{(XR)}_{x_0 x_1|y_1}p_{\rm ss}(x_1,y_1)-W^{(XR)}_{x_1 x_0|y_1}p_{\rm ss}(x_0,y_1).
\label{thermo_J}
\end{align}
The sum of the rates of the entropy change in reservoirs $XL$ and $XR$ is then expressed as
\begin{align}
\sigma^X_r :=J_X F_X.
\label{ep_x}
\end{align}
We can also define $F_Y$, $J_Y$, and $\sigma^Y_r$ in the same manner.

We next consider the concepts of information flow and information affinity~\cite{HorowitzEsposito}.
We first introduce the stochastic mutual information~\cite{SagawaUeda2} between $X$ and $Y$:
\begin{equation}
I(x;y):={\rm ln}p_{\rm ss}(x,y)-{\rm ln}[p_{\rm ss} (x)p_{\rm ss}(y)],
\end{equation}
whose ensemble average is the mutual information in the steady state.
Here, $p_{\rm ss}(x)=\sum_{y}p_{\rm ss}(x,y)$ represents the marginal distribution
of $X$, and $p_{\rm ss}(y)$ is that for $Y$.
The information affinity is defined as
\begin{align}
 F_I &:=I(x_1;y_0)-I(x_0;y_0)+I(x_0;y_1)-I(x_1;y_1) \\
 &={\rm ln}\frac{p_{\rm ss}(x_0, y_1)p_{\rm ss}(x_1, y_0)}{p_{\rm ss}(x_0, y_0)p_{\rm ss}(x_1, y_1)}.
 \label{info_affinity}
 \end{align}
Correspondingly, the probability current that is conjugate with $F_I$ is defined as
\begin{align}
J_I:=&(W^{(YL)}_{y_1 y_0|x_0}+W^{(YR)}_{y_1 y_0|x_0})p_{\rm ss}(x_0,y_0)\nonumber\\
&-(W^{(YL)}_{y_0 y_1|x_0}+W^{(YR)}_{y_0 y_1|x_0})p_{\rm ss}(x_0,y_1).
\label{infoflow}
\end{align}
Using Eqs.~(\ref{info_affinity}) and~(\ref{infoflow}), the information flow is defined as\begin{equation}
\mathcal{I}:=J_I F_I,
\end{equation}
which represents the change rate in the mutual information induced by transitions in $X$.

We now consider the generalized second law of thermodynamics for the four-state model.
First, the conventional second law for the entire system is given by
\begin{equation}
\sigma := \sigma^X_r + \sigma^Y_r \geq 0,
\label{total_second}
\end{equation}
which implies that the total entropy production written as $\sigma$ is nonnegative.
In Ref.~\cite{HorowitzEsposito}, it has been shown that the second law~(\ref{total_second}) can be decomposed into two inequalities that constitute the generalized second law. We introduce the partial entropy productions~\cite{ShiraishiSagawa} associated with $X$ and $Y$ as
\begin{align}
\sigma^X &:=\sigma^X_r+\mathcal{I} = J_X F_X+J_I F_I, \\ 
\sigma^Y&:=\sigma^Y_r-\mathcal{I} = J_Y F_Y - J_I F_I,
\end{align}
which make the decomposition of the total entropy production:
\begin{equation}
\sigma^X + \sigma^Y = (\sigma^X_r+\mathcal{I}) + (\sigma^Y_r-\mathcal{I}) = \sigma.
\end{equation}
The generalized second law states that the partial entropy productions are nonnegative individually:
\begin{equation}
\sigma^X \geq 0 , \ \sigma^Y \geq 0.
\label{infothermoX}
\end{equation}
Both of these inequalities are stronger than the conventional second law~(\ref{total_second}).
In the generalized second law~(\ref{infothermoX}), the conventional thermodynamic entropy production~(\ref{ep_x}) and the information flow are treated on an equal footing. Therefore, inequality~(\ref{infothermoX}) can be regarded as the second law of information thermodynamics for continuous information processing.

Inequality~(\ref{infothermoX}) implies that the entropy change in the reservoirs, $\sigma^X_r$ and $\sigma^Y_r$, are bounded by the informational flow $ - \mathcal I$ and $\mathcal I$, respectively.
For example, in the case of $\mathcal{I}>0$ (i.e., $Y$ plays the role of the demon), the conventional entropy change of $X$
 can be negative up to $-\mathcal{I}$. In contrast, $Y$ needs an additional cost of at least $\mathcal{I}$ for the continuous measurement and
 feedback control.

\subsection{\label{onsager coefficient}Onsager coefficient and reciprocity}
We now discuss our main results for the four-state model. 
First, we can show that if all of the affinities including the information affinities are zero,
all of the currents are zero:
 \begin{equation}
F_X=0, F_I=0\Rightarrow J_X=0, J_I=0.
 \label{zero order condition}
 \end{equation}
We will prove this in Appendix A in a more general setup.

As a consequence of  (\ref{zero order condition}), we find that our linear irreversible thermodynamics is applicable if only $F_X$ and $F_I$ are nearly zero, even when the entire system is far from equilibrium as $F_Y$ is not close to zero. 
In other words, our framework is applicable if only the engine is close to equilibrium both in terms of the thermodynamic and information affinities, even when the demon itself is far from equilibrium.
This is a crucial characteristic of our formulation, which is contrastive to  conventional linear irreversible thermodynamics, where the entire system must be close to equilibrium.
We emphasize that this characteristic is not a result of the time-scale separation.  Even if the engine and the demon have the same time scale and they interact with each other, we can apply linear irreversible thermodynamics only to the engine by taking into account the information affinity.
%This is a desirable feature for the linear theory of information thermodynamics.

Let us discuss the above point more quantitatively.  Even if $F_Y \neq 0$, we can make $F_X = 0$ and $F_I = 0$ by appropriately choosing the parameters in the transition rates (\ref{bipartite1_1}).
In fact, as explicitly shown in Appendix B, if we assume that $\Gamma := \Gamma^{(\nu)}_{y_i}=\Gamma^{(\nu)}_{x_i}$ does not depend on $\nu$, $x_i$, $y_i$, then $F_I = 0$ holds independent of $F_X$ and $F_Y$.  
In such a situation, the energy barrier cannot change depending on the state of the engine or the demon, which means that the demon cannot insert a wall between the engine and the reservoirs.
This intuitive picture is consistent with the absence of the information affinity, $F_I = 0$.
Therefore, we can make $F_X \simeq 0$ and $F_I \simeq 0$ just by making $\mu_{XL}\simeq\mu_{XR}$ and $\Gamma := \Gamma^{(\nu)}_{y_i} \simeq \Gamma^{(\nu)}_{x_i}$, even when $F_Y \not\simeq 0$ (i.e., $\mu_{YL} \not\simeq \mu_{YR}$).

We now define the Onsager coefficients associated with the thermodynamic and informational affinities:
 \begin{align}
 L_{XX}&:=\left. \frac{\partial J_X[F_X, F_I]}{\partial F_X}\right |_{\substack{F_X=0\\F_I=0}},\ 
 L_{XI}:=\left. \frac{\partial J_X[F_X, F_I]}{\partial F_I}\right |_{\substack{F_X=0\\F_I=0}},\nonumber\\
 L_{IX}&:=\left. \frac{\partial J_I[F_X, F_I]}{\partial F_X}\right |_{\substack{F_X=0\\F_I=0}},\ 
 L_{II}:=\left. \frac{\partial J_I[F_X, F_I]}{\partial F_I}\right |_{\substack{F_X=0\\F_I=0}}.
 \label{def Oc}
 \end{align}
 This is indeed well-defined, because condition~(\ref{zero order condition}) ensures that $J_X$ and $J_I$ are single-valued functions of $F_X$ and $F_I$ around
 $F_X=0$ and $F_I=0$.
The currents can then be expanded to the linear order as
 \begin{align}
 J_X&=L_{XX}F_X+L_{XI}F_I,\nonumber\\
 J_I&=L_{IX}F_X+L_{II}F_I.
 \label{linear}
 \end{align}
 The linear expansion~(\ref{linear}) as a consequence of~(\ref{zero order condition}) implies that linear irreversible thermodynamics is well-defined even in the presence of information flow.
 
Since the information affinity is not one of the conventional thermodynamic affinities, we cannot apply the conventional argument in linear irreversible thermodynamics to prove the Onsager reciprocity for the coefficient~(\ref{def Oc}).
However, the Onsager reciprocity indeed holds:
\begin{equation}
 L_{XI}=L_{IX},
 \label{reciprocity}
 \end{equation}
 which is our main results in the case of the four-state model. We will prove Eq.~(\ref{reciprocity}) for a more general setup in Sec.~\ref{generalbipartite}.
The Onsager reciprocity~(\ref{reciprocity}) implies that information thermodynamics has the same structure as conventional thermodynamics in the linear regime.
We again emphasize that this is not a straightforward consequence of~(\ref{infothermoX}). In fact, a standard proof of the Onsager reciprocity based on the fluctuation-dissipation theorem (or equivalently, the fluctuation theorem) does not apply to information thermodynamics. Instead, we need a careful generalization of the Schnakenberg network theory~\cite{Schnakenberg} as discussed in Sec.~\ref{generalbipartite}. 

We note that by using Eqs.~(\ref{linear}) and~(\ref{reciprocity}), the generalized second law $\sigma^X\geq 0$ in~(\ref{infothermoX}) reduces to
\begin{equation}
L_{XX}\geq0, \ L_{II}\geq0,\ L_{XX}L_{II}-L_{XI}^2\geq0.
\label{Onsager_second}
\end{equation}
 
\subsection{Information-thermodynamic efficiency at maximum power\label{ssmaxpower}}

As a special application of Eq.~(\ref{reciprocity}), we derive the universal bound for the efficiency at maximum power of information-driven engines in the linear regime.
The efficiency at maximum power for conventional linear irreversible thermodynamics is known to be the half of the maximum efficiency
in thermodynamics, which has been derived on the basis of the Onsager reciprocity~\cite{Broeck}.
Therefore, it is naturally expected that the information-thermodynamic efficiency at maximum
 power is also the half of the maximum efficiency. In the following, we will show
 that this is indeed the case. 

 We assume that $Y$ plays the role of the demon and $X$ is driven by the information flow, that is,
$J_XF_X<0$. The information-thermodynamic efficiency is then defined as
\begin{equation}
\eta :=-\frac{J_XF_X}{J_IF_I},
\end{equation}
which satisfies $\eta \leq \eta_{\rm max}:=1$ from the first inequality in~(\ref{infothermoX}), and the equality is achievable
 in the strong coupling condition.
The power  is defined as the work extraction per unit time:
\begin{equation}
P : = -J_XF_X.
\end{equation}
We note that, strictly speaking, $P$ is the power multiplied by the inverse temperature.

We now fix $F_I$ and the Onsager coefficients, and then optimize $F_X$ to make $P$ maximum. First, the maximum value of $P$ is achieved if $F_X$ takes the optimal value $F^{\ast}_X$ that is given by
\begin{equation}
F^{\ast}_X :=-\frac{L_{XI}}{2L_{XX}}F_I.
\end{equation}
With $F^{\ast}_X$, the efficiency is calculated as
\begin{equation}
\eta^{\ast}=\frac{q^2}{2(2-q^2)},
\label{effq}
\end{equation}
where $q$ represents the coupling constant defined as
\begin{equation}
q :=\frac{L_{XI}}{\sqrt{L_{XX}L_{II}}}.
\end{equation}
Since the third inequality in~(\ref{Onsager_second}) implies $-1\leq q \leq1$, Eq.~(\ref{effq}) leads to the upper bound
for $\eta^{\ast}$ as
\begin{equation}
\eta^{\ast} \leq \frac{1}{2},
\label{maxeff}
\end{equation}
where the equality is achieved with the strong coupling condition $|q|=1$.
Inequality~(\ref{maxeff}) implies that the information-thermodynamic efficiency at maximum power is given by the half of the maximum efficiency $\eta_{\rm{max}}=1$.

 \section{\label{generalbipartite}GENERAL BIPARTITE SYSTEMS}
 
 We next consider linear irreversible thermodynamics of information-driven engines in the general setup of bipartite Markov jump processes.
 The argument in the previous section is regarded as a special case of that in the present section.
Our formulation is based on the Schnakenberg network theory~\cite{Schnakenberg}.

\subsection{\label{setupG}Setup}
We consider a bipartite Markov jump process of system $Z$ that consists of two subsystems $X$ and $Y$. States of $Z$ are labeled by $z=(x,y)$. The probability distribution of $z$ at time $t$ is denoted by $p(z,t)$, which obeys the master equation in the same form as Eq.~(\ref{mastereq}).
We consider the steady state of $Z$, where the probability distribution is written in a time-independent form $p(z,t)=:p_{\rm ss}(z)$.
The transition rates are assumed to satisfy the bipartite condition:
$W^{(\nu)}_{z z'} =0$ holds if both of $x\neq x'$ and $y\neq y'$ are satisfied,
which implies that $x$ and $y$ do not change simultaneously.
We also assume that 
\begin{equation}
W^{(\nu)}_{zz'}\neq 0\Leftrightarrow W^{(\nu)}_{z'z}\neq 0,
\label{finite}
\end{equation}
which implies that for any transition a backward transition always exists.

To analyze the thermodynamic properties of this system, we introduce a graph representation of the dynamics.
The graph $\mathcal{G}:=(\mathcal{V},\mathcal{E})$ consists of the set $\mathcal{V}$ of vertices and the set $\mathcal{E}$ of directed edges.
 A vertex in the graph represents a state of $Z$. 
A directed edge represents forward and backward transitions between two states of $Z$ induced by reservoir $\nu$ with nonzero transition rate.
In other words, a single directed edges $e:=(z'\xrightarrow{\nu}z)$ corresponds to a pair of transitions $(W^{(\nu)}_{zz'}, W^{(\nu)}_{z'z})$ 
with $W^{(\nu)}_{zz'}\neq 0$ and $W^{(\nu)}_{z'z}\neq 0$. We note that we can arbitrarily choose the direction of an edge.
For example, the state space of the four-state model (Fig.\ref{4state}(b)) is described as Fig.\ref{4stategraph} in the directed graph representation.
Under this assignment,
$(z'\xrightarrow{\nu}z)\in \mathcal{E}\Rightarrow(z\xrightarrow{\nu}z')\notin \mathcal{E}$ is satisfied.
Because the system is bipartite, each edge describes a transition in $X$ or $Y$. A transition in $X$ is described as $((x',y)\xrightarrow{\nu}(x,y))$ ($x \neq x'$), while a transition in $Y$ is described as $((x,y')\xrightarrow{\nu}(x,y))$ ($y\neq y'$).
Correspondingly, we divide the set of all edges into two sets; one is the set of transitions in $X$, and the other is those in $Y$:
\begin{align}
\mathcal{E}^X &:=\{e=((x',y)\xrightarrow{\nu}(x,y))\in \mathcal{E}\}\ (x \neq x'),\nonumber\\
\mathcal{E}^Y &:=\{e=((x,y')\xrightarrow{\nu}(x,y))\in \mathcal{E}\}\ (y\neq y').
\end{align}

We here consider the cycle decomposition~\cite{Schnakenberg} to analyze this system.
 We first introduce a directed cycle $C$ as a directed sequence of connected edges with the same initial and terminal vertex in $\mathcal{G}$: $C=(z_1\xrightarrow{\nu_1}z_2\dots\xrightarrow{\nu_n}z_1)$, where either $(z_i\xrightarrow{\nu_i}z_{i+1})\in \mathcal{E}$ or $(z_{i+1}\xrightarrow{\nu_i}z_{i})\in \mathcal{E}$ holds for any $i$ with identifying $z_{n+1}$ with $z_{1}$.
We then define a cycle basis $\mathcal{C}=\{C_1,C_2,\cdots,C_s\}$ as a set of directed cycles such that any other cycles in $\mathcal{G}$ can be expressed by a linear combination of cycles in $\mathcal{C}$~\cite{Graph}. 
Although a cycle basis is not unique, the number of the cycles in a cycle basis is always given by $s:=|\mathcal{E}|-| \mathcal{V} |+1$, where $|\mathcal{E}|$ is the 
 number of edges and $| \mathcal{V}|$ is the number of vertices.
We also note that a cycle basis is not necessarily a fundamental cycle basis in our argument, in contrast to the original argument
 by Schnakenberg~\cite{Schnakenberg}.
 
 For edge $e \in \mathcal{E}$, we define the edge affinity and edge current as
\begin{align}
 F_e:=&{\rm ln}\frac{W_{zz'}^{(\nu)}}{W_{z'z}^{(\nu)}},\\
\label{thermo_affinity_b}
J_e:=&W_{zz'}^{(\nu)}p_{\rm ss}(z')-W_{z'z}^{(\nu)}p_{\rm ss}(z),
\end{align}
where $F_e$ is finite due to condition~(\ref{finite}).
The total entropy production in $Z$ is then given by
\begin{equation}
\sigma :=\sum_{e\in \mathcal{E}} J_e F_e.
\end{equation}

We define the backward transition edge of $e =(z'\xrightarrow{\nu}z)$ as $e^{\dagger}:=(z\xrightarrow{\nu}z')$.
Following the argument by Schnakenberg~\cite{Schnakenberg}, we define a $|\mathcal E| \times s$ cycle matrix $S$ by
\begin{equation}
S(e, C_k) :=
\begin{cases}
1 &({\rm if} \ e\in C_k),\\
-1 &({\rm if} \ e^{\dagger} \in C_k),\\
0 &({\rm otherwise}),
\end{cases}
\end{equation}
where $S(e, C_k)$ is the matrix element of $S$ with $e \in \mathcal E$ and $C_k \in \mathcal{C}$. Here, $e \in C_k$ means that $e$ is one of the edges in $C_k$.
 We then assign the affinity $F(C_k)$ to cycle $C_k$ by
\begin{equation}
F(C_k):=\sum_{e \in \mathcal{E}}S(e, C_k)F_e,
\label{currentdef1}
\end{equation}
and assign the current $J(C_k)$ to cycle $C_k$ as the solution of
\begin{equation}
J_e=\sum_{C_k \in \mathcal{C}} S(e, C_k) J(C_k).
\label{currentdef2}
\end{equation}
We note that the cycle matrix $S$ is full rank~\cite{Graph}, and therefore $J(C_k)$ always exists and is uniquely determined by Eq.~(\ref{currentdef2}). 
By using the above notations, we can express the total entropy production as~\cite{Schnakenberg}
\begin{align}
 \sigma&=\sum_{e\in \mathcal{E}} J_e F_e \\
 &=\sum_{e\in \mathcal E}\sum_{C_k \in \mathcal{C}}S(e,C_k)J(C_k)F_e \\
 &=\sum_{C_k \in \mathcal{C}}J(C_k)F(C_k).
\end{align}

 We next define the effective affinity of edge $e =(z'\xrightarrow{\nu}z)$~\cite{Schnakenberg}:
 \begin{equation}
 \mathcal{F}_e:=F_e+{\rm ln}\frac{p_{\rm ss}(z')}{p_{\rm ss}(z)}.
 \label{effa}
 \end{equation}
The effective affinity of edge $e \in \mathcal E$ is regarded as the conjugate of the current of $e$, due to the fact that $\mathcal{F}_e=0$
holds if and only if $J_e=0$. 
In contrast, we refer to $F_e$ as a bare thermodynamic affinity.
Using the definition~(\ref{currentdef1}) and the cyclic property, we obtain
 \begin{equation}
 F(C_k) = \sum_{e\in \mathcal E} S(e,C_k)\mathcal{F}_e.
 \label{FCk}
 \end{equation}
 The thermodynamic properties of the system are expressed in terms of affinities and currents of a cycle basis.

 \subsection{\label{partial affinity}Partial affinity and partial entropy production}
 
 We next introduce important concepts to formulate linear irreversible thermodynamics with information flow in line with Ref.~\cite{HorowitzEsposito}.
 First, we define the partial affinities of cycles associated with $X$ by
 \begin{equation}
 \mathcal{F}^X(C_k):=\sum_{e \in \mathcal E^X}S(e,C_k) \mathcal{F}_e. \label{affinity}
 \end{equation}
We remark that the partial affinity is different from the cycle affinity, because the summation is not taken over $\mathcal E$ but only over $\mathcal E^X$. 
We then define the partial entropy production associated with $X$ by
 \begin{equation}
 \sigma^X=\sum_{C_k \in \mathcal{C}}J(C_k)\mathcal{F}^X(C_k),
 \label{cycleX}
 \end{equation}
 which satisfies the generalized second law~\cite{HorowitzEsposito}:
 \begin{equation}
 \sigma^X \geq 0.
 \label{bipartite_second}
 \end{equation}

 We next classify cycles in a cycle basis into the following three classes: a local cycle of $X$, a local cycle of $Y$, and a global cycle.
A local cycle of $X$ is defined as a cycle which consists only of edges in $X$, like $\{(x_1, y)\xrightarrow{\nu_1}(x_2, y)\xrightarrow{\nu_2} \cdots \xrightarrow{\nu_n}(x_1, y)\}$.
Let $\mathcal C^X := \{ C_1^X, C_2^X, \cdots, C_n^X \}$ be the set of local cycles of $X$ in $\mathcal C$, which has $n$ cycles. We define $\mathcal C^Y := \{ C^Y_1, \cdots, C^Y_m \}$ in the same manner, which has $m$ cycles.
A global cycle is defined as a cycle that consists of edges in both $X$ and $Y$. Let $\mathcal C^G := \{ C^G_1, \cdots, C^G_l\}$ be the set of global cycles in $\mathcal C$, which has $l$ cycles.
We then have $\mathcal C = \mathcal C^X \cup \mathcal C^Y \cup \mathcal C^G$ by definition.
We note that $n+m+l=|\mathcal E|-| \mathcal V|+1$ holds for any cycle basis, while each of $n$, $m$, $l$ depends on the choice of a cycle basis.
In the following, we choose a cycle basis such that $m$ ($=| \mathcal C^Y|$) takes the maximum value for a given graph (see Appendix A for details).

With the above classification of cycles, we have the following properties of the affinities. We first consider a local cycle $C^X_k \in \mathcal C^X$.
By noting that $S(e,C_k^X) = 0$ for $e \not\in \mathcal E^X$, we have from Eq.~(\ref{FCk})
\begin{equation}
\mathcal{F}^X(C^X_k)=F(C^X_k),
\end{equation} 
which implies that the partial affinity of $X$ for a local cycle of $X$ is given by the sum of the bare thermodynamic affinities. We next consider a local cycle $C^Y_k \in \mathcal C^Y$.
By noting that $S(e,C_k^Y) = 0$ for $e \in \mathcal E^X$ and $C_k^Y \in \mathcal C^Y$, we have
\begin{equation}
\mathcal{F}^X(C^Y_k)=0,
\end{equation} 
which implies that the partial affinity of $X$ for a local cycle of $Y$ vanishes.
For a global cycle $C^G_k \in \mathcal C^G$, we have
\begin{equation}
\mathcal{F}^X(C^G_k) = F^X(C^G_k) + \mathcal{F}^I({C_k^G}),
\label{C_G}
\end{equation}
where the first term on the right-hand side is defined as 
\begin{equation}
F^X(C^G_k):=\sum_{e \in \mathcal{E}^X}S(e, C^G_k)F_e,
\end{equation}
and the second term is the information affinity~\cite{HorowitzEsposito}:
 \begin{align}
\mathcal{F}^I({C_k^G}) :=&\sum_{e\in \mathcal E^X} S(e,C_k^G){\rm ln}\frac{p_{\rm ss}(z')}{p_{\rm ss}(z)}\\
=&\sum_{e\in \mathcal E^X} S(e,C_k^G)(I(x';y)-I(x;y)).
 \end{align}
In the second line above, we used
 \begin{equation}
 \sum_{e\in \mathcal E^X} S(e,C_k^G){\rm ln}\frac{p_{\rm ss}(x')}{p_{\rm ss}(x)}=0.
 \end{equation}
The information affinity describes the change rate in the mutual information in the edges of $C_k$ associated with $X$.
In contrast, $F^X(C^G_k)$ includes the bare thermodynamic affinities (i.e., $F_e$). We note that our theory is applicable to situations where $X$ and $Y$ exchange energy (i.e. $F^X(C^G_k) \neq 0$).

The partial entropy production~(\ref{cycleX}) is then rewritten as
 \begin{equation}
 \sigma^X=\sum_{C_k^X \in \mathcal{C}^X} J(C_k^X)F(C_k^X) + \sum_{C_k^G \in \mathcal{C}^G}J(C_k^G)\mathcal{F}^X(C_k^G).
 \label{cycleX2}
 \end{equation}
 We note that if $ F^X(C^G_k) = 0$ for any $C^G_k \in \mathcal C^G$, the second term on the right-hand side of~(\ref{cycleX2}) only describes the information flow:
\begin{equation}
\mathcal I := \sum_{C_k^G \in \mathcal{C}^G}J(C_k^G)\mathcal{F}^I(C_k^G).
\end{equation}

\begin {figure}[H]
\centering
\includegraphics[width=8cm]{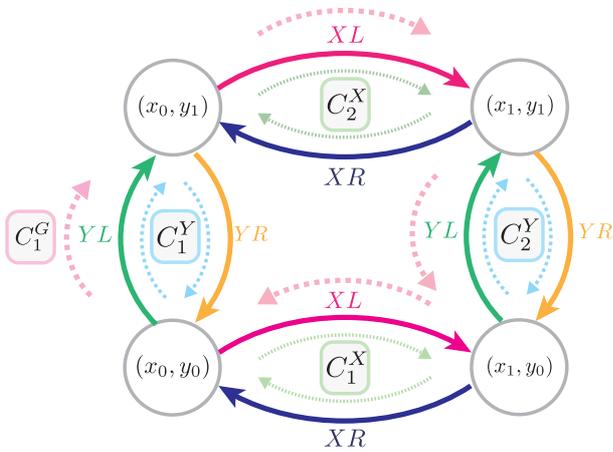}
\caption{(Color online) A directed graph of the four-state model and its cycle basis. The solid lines represent two transitions between two state, and the dotted lines represent edges of a cycle basis. Note that we can arbitrarily choose the direction of each edge.}
\label{4stategraph}
\end{figure}

As a simple example, we revisit the four-state model discussed in Sec.~\ref{paradigmatic}. 
Figure~\ref{4stategraph} shows a cycle basis of the four-state model. There are five cycles in the cycle basis, which are classified into the local and global cycles as
 \begin{align}
 C^X_1:=&\{\!(x_0,\! y_0)\!\xrightarrow{XL}\!(x_1,\! y_0)\!\xrightarrow{XR}\!(x_0,\! y_0)\},\nonumber\\
 C^X_2:=&\{\!(x_0,\! y_1)\!\xrightarrow{XL}\!(x_1,\! y_1)\!\xrightarrow{XR}\!(x_0,\! y_1)\},\nonumber\\
 C^Y_1:=&\{\!(x_0,\! y_0)\!\xrightarrow{YL}\!(x_0,\! y_1)\!\xrightarrow{YR}\!(x_0,\! y_0)\},\nonumber\\
 C^Y_2:=&\{\!(x_1,\! y_0)\!\xrightarrow{YL}\!(x_1,\! y_1)\!\xrightarrow{YR}\!(x_1,\! y_0)\},\nonumber\\
 C^G_1:=&\{\!(x_0,\! y_0)\!\xrightarrow{YL}\!(x_0,\! y_1)\!\xrightarrow{XL}\!(x_1,\! y_1)\!\nonumber\\
 &\xrightarrow{YL}\!(x_1,\! y_0)\!\xrightarrow{XL}\!(x_0,\! y_0)\!\}.
 \end{align}
 Therefore, the partial entropy production~(\ref{cycleX2}) reduces to 
 \begin{equation}
\sigma^X = J(C^X_1)\mathcal{F}^X(C^X_1)\!+\!J(C^X_2)\mathcal{F}^X(C^X_2)\!+\!J(C^G_1)\mathcal{F}^X(C^G_1),
\label{a1}
\end{equation}
 where
\begin{align}
J(C^X_1):= &W^{(XR)}_{x_0 x_1|y_0}p_{\rm ss}(x_1,y_0)-W^{(XR)}_{x_1 x_0|y_0}p_{\rm ss}(x_0,y_0),\nonumber\\
J(C^X_2):=&W^{(XR)}_{x_0 x_1|y_1}p_{\rm ss}(x_1,y_1)-W^{(XR)}_{x_1 x_0|y_1}p_{\rm ss}(x_0,y_1),\nonumber\\
J(C^G_1):=&(W^{(YL)}_{y_1 y_0|x_0}+W^{(YR)}_{y_1 y_0|x_0})p_{\rm ss}(x_0,y_0)\nonumber\\
&-(W^{(YL)}_{y_0 y_1|x_0}+W^{(YR)}_{y_0 y_1|x_0})p_{\rm ss}(x_0,y_1),\nonumber\\
\mathcal{F}^X(C^X_1)= &\mathcal{F}^X(C^X_2):= \beta (\mu_{XL}-\mu_{XR}) ,\nonumber\\
\mathcal{F}^X(C^G_1):= &{\rm ln}\frac{p_{\rm ss}(x_0, y_1)p_{\rm ss}(x_1, y_0)}{p_{\rm ss}(x_0, y_0)p_{\rm ss}(x_1, y_1)}.
\label{a2}
\end{align}
Comparing Eqs.~(\ref{a2}) with Eqs.~(\ref{thermo_affinity}),~(\ref{thermo_J}),~(\ref{info_affinity}),~(\ref{infoflow}), we 
obtain
\begin{align}
J_X &= J(C^X_1) + J(C^X_2),\nonumber\\
J_I &= J(C^G_1),\nonumber\\
F_X &= \mathcal{F}^X(C^X_1)=\mathcal{F}^X(C^X_2),\nonumber\\
F_I &= \mathcal{F}^X(C^G_1).
\end{align}
Inequality~(\ref{bipartite_second}) reduces to the first inequality of~(\ref{infothermoX}) in the four-state model.

\subsection{\label{ORRinG}Onsager reciprocity for the general setup}

We now discuss our main results for general bipartite systems. 
We first abbreviate
\begin{align}
\mathcal{F}^X_k :=\mathcal{F}^X(C_k),\nonumber\\
\mathcal{J}^X_k :=J^X(C_k),
\end{align}
where we defined $N_X:=\{1, \cdots, n\}$ and $N_G:=\{n+1, \cdots, n+l\}$ with $N:=N_X\cup N_G$, and
labeled cycles as
\begin{equation}
 C_k:=
\begin{cases}
C^X_k\ &k\in N_X\\
C^G_{k-n}\ &k\in N_G.
 \end{cases}
\end{equation}

As shown in Appendix A, we have
\begin{equation}
\forall k\in N, \ \mathcal{F}^X_k=0 \ \ \Rightarrow \ \ \forall k\in N, \ \mathcal{J}^X_k=0,
\label{zeroG}
\end{equation}
which is a general form of condition~(\ref{zero order condition}).
We emphasize that condition~(\ref{zeroG}) is not a consequence of the conventional linear irreversible thermodynamics, because $\mathcal{F}_k^X$ are not affinities of the cycle basis. 
As is the case for the four-state model, condition~(\ref{zero order condition}) implies that our linear irreversible thermodynamics is applicable  if only $\mathcal{F}^X_k \simeq 0$ for any $k\in N$, even when the entire system is far from equilibrium.

Condition~(\ref{zeroG}) ensures the validity of the linear expansion
\begin{equation}
\mathcal{J}^X_i = \sum_{j=1}^{n+l} L_{ij}^X \mathcal{F}^X_j
\end{equation}
 for $i\in N$. Therefore, the Onsager coefficient is well-defined:
\begin{equation}
L_{ij}^X := \left.\frac{\partial \mathcal{J}^X_i}{\partial \mathcal{F}^X_j}\right|_{\forall k, \mathcal{F}^X_k=0},
\end{equation}
which implies that linear irreversible thermodynamics with information affinities is a consistent framework.
We note that for $i \in N_X$ and $j \in N_G$, $L_{ij}$ describes the driving of a conventional thermodynamic current by the information affinity, while for $i\in N_G$ and $j \in N_X$, $L_{ij}$ describes the driving of an information current by the conventional affinity.

We can then show the Onsager reciprocity:
\begin{equation}
L_{ij}^X = L_{ji}^X,
\label{reciprocity_G}
\end{equation}
which is the main result of this paper in the general setup. We will prove
Eq.~(\ref{reciprocity_G}) in the next subsection. Equality~(\ref{reciprocity}) in Sec.~\ref{paradigmatic} is a special case of Eq.~(\ref{reciprocity_G}).
For $i \in N_X$ and $j\in N_G$, or for $i\in N_G$ and $j \in N_X$, the Onsager reciprocity represents a novel symmetry between the thermodynamic and informational quantities.

\subsection{\label{proof}Proof of the main result}
 In this subsection, we prove our main result~(\ref{reciprocity_G}). 
Before that, we briefly mention the outline of the proof of the zero currents condition~(\ref{zeroG}), while the complete proof is shown in Appendix A. If all affinities of cycles in the cycle basis are zero, then all currents of the cycles are zero, because $S$ is full rank.
However, this fact does not directly lead to condition~(\ref{zeroG}), because $\mathcal{F}_k^X$ are not affinities of the cycle basis.
To prove condition~(\ref{zeroG}), we need to use the graph contraction method (see Appendix A for details). Roughly speaking, we can construct a new Markov
jump system whose cycle affinities are $\mathcal{F}_k^X$ and whose currents are $\mathcal{J}^X_k$.
 
We now prove Eq.~(\ref{reciprocity_G}) by assuming condition~(\ref{zeroG}).
 From condition~(\ref{zeroG}), $\mathcal{J}^X_i$ can be expanded with $\mathcal{F}^X_j$:
 \begin{equation}
\mathcal{J}^X_i=\sum_{j}L_{ij}\mathcal{F}^X_j+O((\mathcal{F}^X_j)^2).
\label{linearJX}
\end{equation}
Since $S(e, C_k)=0$ holds for $e\in\mathcal{E}^X$ and $C_k \in \mathcal{C}^Y$, ${J}_e$ with $e\in\mathcal{E}^X$ can be expressed as a linear combination of $\mathcal{J}^X_k$:
 \begin{equation}
{J}_e=\sum_{k \in N}S(e, C_k) \mathcal{J}^X_k.
\label{linearJe}
\end{equation}
We note that from Eqs.~(\ref{linearJX}) and~(\ref{linearJe}), the linear regime in
terms of the partial affinities (i.e., $\mathcal{F}^X_k\simeq 0$, $\forall k \in N$) is equivalent to the linear regime in terms of the edge currents (i.e., ${J}_e\simeq 0$, $ \forall e\in\mathcal{E}^X$).
We next expand $\mathcal{F}_e$ to the linear order of $J_e$ as
\begin{align}
\mathcal{F}_e&={\rm ln}\frac{W^{(\nu)}_{zz'}p_{\rm ss}(z')}{W^{(\nu)}_{z'z}p_{\rm ss}(z)} \\
&={\rm ln}\left[1+\frac{W^{(\nu)}_{zz'}p_{\rm ss}(z')-W^{(\nu)}_{z'z}p_{\rm ss}(z)}{W^{(\nu)}_{z'z}p_{\rm ss}(z)}\right]\\
&={\rm ln}\left[1+\frac{J_e}{{\alpha_e}^{-1}+O(J_e)}\right]\\
&=\alpha_e J_e+O(J_e^2)\label{linearFe},
\end{align}
where we defined
\begin{equation}
 \alpha_e:=\left. \frac{1}{W^{(\nu)}_{z'z}p_{\rm ss}(z)}\right|_{J_e=0}=\left. \frac{1}{W^{(\nu)}_{zz'}p_{\rm ss}(z')}\right|_{J_e=0},
 \end{equation}
and assumed that $0<\alpha_e<\infty$.
Using linear expansions~(\ref{linearJe}) and~(\ref{linearFe}), the affinities in $X$ are calculated as
\begin{align}
\mathcal{F}^X_i&=\sum_{e\in \mathcal E^X} S(e, C_i)\mathcal{F}_e\\
			&=\sum_{e\in \mathcal E^X} S(e, C_i)\alpha_e{J}_e+O(J_e^2)\\ 
			&=\sum_{e\in \mathcal E^X}\sum_{j} S(e, C_i)\alpha_e S(e, C_j) \mathcal{J}^{X}_j+O((\mathcal{J}^X_j)^2).
			\label{linearFX}
\end{align}
By defining a $(s-m)\times(s-m)$ matrix $M$ as
\begin{equation}
M_{ij}=\sum_{e\in \mathcal E^X} S(e, C_i)\alpha_e S(e, C_j),
\label{defM}
\end{equation}
and insulting~(\ref{linearJX})
into Eq.~(\ref{linearFX}), we find
\begin{align}
\mathcal{F}^X_i=\sum_{j,k}M_{ij}L_{jk}\mathcal{F}^X_k+O((\mathcal{F}^X_k)^2).
\end{align}
Since $\mathcal{F}^X_i$ is arbitrary, $M$ is given by the inverse matrix of $L$: \begin{align}
L=M^{-1}.
\end{align}
The form of~(\ref{defM}) implies that $M$ is a symmetric matrix $M_{ij}=M_{ji}$.
This proves that the Onsager matrix $L$ is also symmetric,
which leads to the Onsager reciprocity~(\ref{reciprocity_G}).

Although the above proof of the Onsager reciprocity is apparently similar to the original argument of Schnackenberg~\cite{Schnakenberg}, 
here we essentially invoked condition~(\ref{zeroG}). 
We need to introduce the graph contraction method to prove condition~(\ref{zeroG}), as described in Appendix A.
In contrast, the graph contraction is not needed in previous works with information reservoirs ~\cite{BaratoSeifert2, BaratoSeifert3},  which again ensures that our work is fundamentally different from previous work.

%%%%%%%%%%%%%%%%%%%%%%%%%%%%%%%%%%%%%%%%%%%%%%%%%%%%%%%%%%%%%%%%%%%%%

\section{\label{conclusion}CONCLUSION}

In this paper, we have developed the framework of linear irreversible thermodynamics for Markov jump systems
with continuous information flow. We have shown that the information affinity and the information current play equivalent roles to conventional thermodynamic affinities and currents in linear irreversible thermodynamics. 
Our main result is the Onsager reciprocity~(\ref{reciprocity_G}) with the information affinity.

As a characteristic of our formulation, it is applicable even when the entire system is far from equilibrium.
Provided only that the thermodynamic affinities and the information affinities of the driven engine are close to zero, linear irreversible thermodynamics applies to the engine, without looking at the demon that can be far from equilibrium.
This is ensured by the fact that all of the information affinities and thermodynamic affinities are
zero, if and only if all of the conjugate currents are zero, as represented by~(\ref{zeroG}).

In conventional statistical mechanics, the Onsager reciprocity is a straightforward consequence of the fluctuation-dissipation theorem (or the Green-Kubo formula), as the equilibrium correlation function is symmetric.
In the modern language, the fluctuation-dissipation theorem can be obtained from the second cumulant of the fluctuation theorem,  and the higher-order generalization of the fluctuation-dissipation theorem can systematically be obtained from the fluctuation theorem~\cite{Andrieux}. 
In contrast, the fluctuation-dissipation theorem is not valid in the presence of information affinity.
Nevertheless, the Onsager reciprocity with information affinity~(\ref{reciprocity_G}) is valid, which implies that our Onsager reciprocity is fundamentally different from the conventional one.

The reason why the fluctuation-dissipation theorem is not valid in the presence of the information affinity is the following.  The average cycle currents of the contracted system introduced in Appendix A are the same as the average cycle currents of the original currents.  However, the higher order cumulants of currents, such as the variance of currents, are different in general.  Therefore, the fluctuation-dissipation theorem is not valid for the contracted graph.
% Therefore, the relations in the conventional linear response theory, such as the
% fluctuation-dissipation theorem, do not hold in our setup,
% except for the Onsager reciprocity.
From the same reason, the Onsager reciprocity cannot be generalized to
 the nonlinear response coefficients, unlike the conventional setup without information.
 
The Onsager reciprocity is a powerful tool to investigate nonequilibrium phenomena,
which are often found in the linear regime.
Therefore, our results would serve a variety
of researches of autonomous information processing.

\begin{acknowledgments}
S.I. is supported by Grant-in-Aid for JSPS Fellows No.
JP15J07404 and JSPS KAKENHI Grant No. JP16K17780.
N.S. is supported by Grant-in-Aid for JSPS Fellows No.
JP14J07602. T.S. is supported by JSPS KAKENHI Grants
No. JP16H02211 and No. JP25103003, and by the Platform for
Dynamic Approaches to Living System of the Japan Agency
for Medical Research and Development (AMED).
\end{acknowledgments}

\appendix

\section{Graph contraction and the proof of~(\ref{zeroG})\label{App}}
To prove condition~(\ref{zeroG}),
 we introduce the contraction of a graph~\cite{Graph}.
 Let us consider a directed graph $\mathcal{G}=(\mathcal{V}, \mathcal{E})$ with the same notation as in Sec.~\ref{generalbipartite}.
The contraction of edge $e=(z_k\xrightarrow{\nu'}z_l) \in \mathcal{E}$ with $k \neq l$ is
defined as follows. We first remove
 edge $e$, and identify its vertices $z_k$, $z_l$ to a new vertex $z_m$, and then construct a new graph $\mathcal{G}/e:=(\tilde{\mathcal{V}},
 \tilde{\mathcal{E}})$ such that
 \begin{equation}
 \tilde{\mathcal{V}}:= \{\tilde z(i) | i \in \mathcal{V} \}
 \end{equation}
 and
\begin{equation}
 \tilde{\mathcal{E}}:=\{(\tilde z(i)\xrightarrow{\nu}\tilde z(j))|(z_i\xrightarrow{\nu}z_j) \in \mathcal{E} \backslash\{e\}\}.
\end{equation}
 Here, $\backslash$ denotes the set difference, and $\tilde z(i)$ represents a map on vertices defined as
 \begin{equation}
 \tilde z(i):=
 \begin{cases}
 z_m&(i=k, l)\\
 z_i&({\rm otherwise}).
\end{cases}
 \end{equation}

Correspondingly, we define the master equation on the contracted graph $\mathcal{G}/e$ as
\begin{equation}
\frac{d}{dt}p(\tilde z(i),t)=\sum_{\tilde z(j), \nu}[\tilde W^{(\nu)}_{\tilde z(i)\tilde z(j)}p(\tilde z(j),t)
-\tilde W^{(\nu)}_{\tilde z(j)\tilde z(i)}p(\tilde z(i),t)],
\label{mastereqc}
\end{equation}
where we defined new transition rates as
\begin{equation}
\tilde W^{(\nu)}_{\tilde z(i) \tilde z(j)}=
\begin{cases}
\frac{p_{ss}(z_j)}{p_{\rm ss}(z_k)+p_{\rm ss}(z_l)}W^{(\nu)}_{z_iz_j}\ &(\tilde z(j)=m)\\
W^{(\nu)}_{z_iz_j}\ &({\rm otherwise}).
\label{change}
\end{cases}
\end{equation}
Note that the contraction can make a loop (i.e., transition from a state to itself), and in such a case, we distinguish two directions of the loop as
$\tilde W_{z_mz_m+}^{(\nu)}:=\tilde W_{\tilde z(i)\tilde z(j)}^{(\nu)}$ and
$\tilde W_{z_mz_m-}^{(\nu)}:=\tilde W_{\tilde z(j)\tilde z(i)}^{(\nu)}$,
while these transition rates play no role in Eq.~(\ref{mastereqc}).
It is easy to show that the steady state probability distribution of the contracted Markov jump system satisfies
\begin{equation}
\tilde p_{\rm ss}(\tilde z(i))=
\begin{cases}
p_{\rm ss}(z_k)+p_{\rm ss}(z_l)\ &(i=k, l)\\
p_{\rm ss}(z_i)\ &({\rm otherwise}).
\end{cases}
\end{equation}
The foregoing contraction does not change currents (i.e., $J_{\tilde e'}=J_{e'}$) or effective affinities (i.e., $\mathcal{F}_{\tilde e'}=\mathcal{F}_{e'}$), where 
$\tilde e'=(\tilde z(i)\xrightarrow{\nu}\tilde z(j))$ is an edge in $\tilde {\mathcal{E}}$ corresponding to edge $e'=(z_i\xrightarrow{\nu}z_j) \in \mathcal{E}$.
If we contract a loop (i.e., $e=(z_k\xrightarrow{\nu '}z_k)$), we just eliminate the loop and do not change other edges or the corresponding transition rates, which neither changes currents nor effective
 affinities.

 We now split the set of edges $\mathcal{E}$ into two disjoint subsets: $\mathcal{E}^X$ and $\mathcal{E}^Y$, and
 consider a new graph $\mathcal{G}/\mathcal{E}^Y:=\mathcal{G}/e_1/e_2\cdots/e_{|\mathcal{E}^Y|}$ by recursively contracting all the edges in $\mathcal{E}^Y=
 \{e_1,e_2,\cdots,e_{|\mathcal{E}^Y|}\}$. This operation does not depend on the order of contraction of edges. In the following,
 we construct the cycle basis of the contracted graph. 
 
 To define the contraction of cycle $C$, we introduce a new notation of a cycle: $C=(\mathcal{V}(C), \mathcal{E}(C))$, where $\mathcal{V}(C)$ and $\mathcal{E}(C)$ are respectively the set of vertices and edges of $C$.
By introducing the contracted set of vertices and edges of $C$ defined as
 \begin{equation}
 \tilde{\mathcal{V}}(C):= \{\tilde z(i) | z_i \in \mathcal{V}(C) \}
 \end{equation}
 and
 \begin{equation}
 \tilde{\mathcal{E}}(C):=\{(\tilde z(i)\xrightarrow{\nu}\tilde z(j))|(z_i\xrightarrow{\nu}z_j) \in \mathcal{E}(C)\backslash\{e\}\},
\end{equation}
we define the contracted cycle $C/e:=( \tilde{\mathcal{V}}(C), \tilde{\mathcal{E}}(C))$ on the contracted graph.
We also define $C/\mathcal{E}^Y:=C/e_1/e_2\cdots/e_{|\mathcal{E}^Y|}$ in the same manner as $\mathcal{G}/\mathcal{E}^Y$, 
and this operation does not depend on the order of contraction of edges. Note that the cycle matrices of the original graph and the
contracted graph satisfy $S(\tilde e,C_i/\mathcal{E}^Y)= S(e,C_i)$ for $e\in \mathcal{E}^X$ and $i \in N$.

Let $\mathcal{C}= \mathcal{C}^X \cup \mathcal{C}^G \cup \mathcal{C}^Y$ be a cycle basis of the graph $\mathcal{G}$
, where $|\mathcal{C}^Y|$ takes the maximum for the given graph. For such a cycle basis, 
$\mathcal{C}/\mathcal{E}^Y :=\{C_{1}/\mathcal{E}^Y, \cdots , C_{n+l}/\mathcal{E}^Y \}$ 
is the cycle basis of the contracted graph $\mathcal{G}/\mathcal{E}^Y$.

\begin {figure}[H]
\centering
\includegraphics[width=8cm]{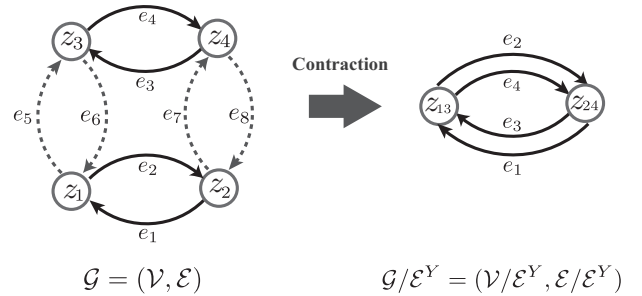}
\caption{The original graph of the four-state model $\mathcal{G}=(\mathcal{V},\mathcal{E})$ and its contracted graph $\mathcal{G}/\mathcal{E}^Y =(\tilde{\mathcal{V}}, \tilde{\mathcal{E}})$, where $\mathcal{E}^Y=\{e_5, e_6, e_7, e_8\}$. }
\label{graphcontraction}
\end{figure}

We show an example of the contraction with the four-state model in Fig.~\ref{graphcontraction}. The original graph is $\mathcal{G} = \{\mathcal{V} , \mathcal{E}\}$, where $\mathcal{V} = \{ z_1, z_2, z_3, z_4 \}$ and $\mathcal{E} = \{ e_1, e_2, e_3, e_4, e_5, e_6, e_7, e_8\}$. The cycle basis of the original graph $\mathcal{C}= \{C^X_1, {C}^X_2, {C}^G_1,C^Y_1,C^Y_2\}$ is
\begin{align}
 {C}^X_1&:=\{\! z_{1} \!\xrightarrow{e_2}\! z_{2} \!\xrightarrow{e_1}\! z_{1} \},\nonumber\\ 
 {C}^X_2&:=\{\! z_{3} \!\xrightarrow{e_4}\! z_{4} \!\xrightarrow{e_3}\! z_{3} \},\nonumber\\
 {C}^G_1&:=\{\! z_{1} \!\xrightarrow{e_5}\! z_{3} \!\xrightarrow{e_4}\! z_{4}
 \!\xrightarrow{e_7^{\dagger}}\! z_{2}\! \xrightarrow{e_2^{\dagger}}\!z_{1}\},\nonumber\\
 {C}^Y_1&:=\{\! z_{1} \!\xrightarrow{e_5}\! z_{3} \!\xrightarrow{e_6}\! z_{1}\},\nonumber\\
 {C}^Y_2&:=\{\! z_{2} \!\xrightarrow{e_7}\! z_{4} \!\xrightarrow{e_8}\! z_{2} \}.
\end{align}
 We split the transition set $\mathcal{E}$ into two disjoint subsets $\mathcal{E}^X := \{ e_1, e_2, e_3, e_4\}$ and $\mathcal{E}^Y := \{ e_5, e_6, e_7, e_8\}$. The contracted graph is then written as $\mathcal{G}/\mathcal{E}^Y = (\mathcal{V}/\mathcal{E}^Y, \mathcal{E}/\mathcal{E}^Y)$ with $\mathcal{V}/\mathcal{E}^Y = \{ z_{13}, z_{24} \}$ and $\mathcal{E}/\mathcal{E}^Y = \mathcal{E}^X = \{ e_1, e_2, e_3, e_4\}$. The cycle basis of the contracted graph is written as $\mathcal{C}/\mathcal{E}^Y= \{C^X_1/\mathcal{E}^Y, {C}^X_2/\mathcal{E}^Y, {C}^G_1/\mathcal{E}^Y\}$ with
\begin{equation}
 \begin{split}
 {C}^X_1/\mathcal{E}^Y=&\{\! z_{13} \!\xrightarrow{e_2}\! z_{24} \!\xrightarrow{e_1}\! z_{13} \},\\
 {C}^X_2/\mathcal{E}^Y=&\{\! z_{13} \!\xrightarrow{e_4}\! z_{24} \!\xrightarrow{e_3}\! z_{13} \},\\
 C^G_1/\mathcal{E}^Y=&\{\! z_{13} \!\xrightarrow{e_4}\! z_{24} \!\xrightarrow{e_2^{\dagger}}\! z_{13} \} .
 \end{split}
 \end{equation}

From the above argument, we obtain the correspondence between the cycle affinities and the currents of the contracted system, and
the partial affinities and the currents of the original system.
First, the cycle affinities of the contracted system are equivalent to the partial affinities of the original 
system:
\begin{align}
F(C_i/\mathcal{E}^Y)&:=\sum_{\tilde e\in \mathcal{E}^X}S(\tilde e,C_i/\mathcal{E}^Y)\mathcal{F}_{\tilde e}\\
&=\sum_{e \in \mathcal{E}^X}\tilde S(e, C_i)\mathcal{F}_{e}\\
&=\mathcal{F}^X(C_i)
\end{align}
for cycle $C_i\in \mathcal{C}^X \cup \mathcal{C}^G$.
Second, the cycle currents of the contracted system are equivalent to those of the original 
system 
\begin{equation}
J(C_i/ \mathcal{E}^Y)=J(C_i),
\end{equation}
 because both of $J(C_i/ \mathcal{E}^Y)$ and $J(C_i)$ are the solutions of
\begin{align}
J_{\tilde e} &=: \sum_{C_i \in \mathcal{C}/\mathcal{E}^Y}S(\tilde e,C_i/\mathcal{E}^Y)J(C_i/ \mathcal{E}^Y),
\end{align}
and
\begin{align}
J_e &= \sum_{C_i \in \mathcal{C}^X \cup \mathcal{C}^G} S(e,C_i)J(C_i),
\end{align}
and both of $J_{\tilde e}=J_e$ and $S(\tilde e,C_i/\mathcal{E}^Y)= S(e,C_i)$ hold for $e \in \mathcal{E}^X$.

Since $\mathcal{C}/\mathcal{E}^Y=\{C_1/\mathcal{E}^Y, \cdots, C_{n+l}/\mathcal{E}^Y\} $ is a cycle basis,
\begin{align}
\forall i \in N, F(C_i/\mathcal{E}^Y)=0\Rightarrow \forall i \in N, J(C_i/\mathcal{E}^Y)=0
\end{align}
is satisfied. This directly leads to condition~(\ref{zeroG}):
\begin{equation}
{\forall i \in N}, \mathcal{F}^{X}(C_i)=0\Rightarrow {\forall i \in N}, J(C_i)=0.
\end{equation}

\section{Steady state distribution of the four-state model}

We explicitly show the steady state distribution of the four-state model.  We consider the master equation (\ref{mastereq}) with the transition rates (\ref{bipartite1}) and (\ref{bipartite1_1}).
We define
\begin{equation}
W_{x_i x_j |y_k} := \sum_\nu W_{x_i x_j |y_k}^{(\nu)}.
\end{equation}
By applying the method in ~\cite{Schnakenberg}, we obtain the steady solution of Eq.~(\ref{mastereq}):
\begin{equation}
p_{\rm{ss}}(x_i, y_i)=\frac{q(x_i, y_i)}{\sum_{x,y} q(x,y)},
\end{equation}
where
\begin{align}
q(x_0,y_0):=&W_{x_0 x_1 |y_0} W_{y_0 y_1 |x_1} (W_{x_1 x_0 |y_1}+W_{y_0 y_1 |x_0})\nonumber\\
&+W_{y_0 y_1 |x_0} W_{x_0 x_1 |y_1} (W_{x_0 x_1 |y_0}+W_{y_1 y_0 |x_1}),\nonumber\\
q(x_1,y_0):= &W_{x_1 x_0 |y_0} W_{y_0 y_1 |x_0} (W_{x_0 x_1 |y_1}+W_{y_0 y_1 |x_1})\nonumber\\
&+W_{y_0 y_1 |x_1} W_{x_1 x_0 |y_1} (W_{x_1 x_0 |y_0}+W_{y_1 y_0 |x_0}),\nonumber\\
q(x_0,y_1):=&W_{y_1 y_0 |x_0} W_{x_0 x_1 |y_0} (W_{x_0 x_1 |y_1}+W_{y_0 y_1 |x_1})\nonumber\\
&+W_{x_0 x_1 |y_1} W_{y_1 y_0 |x_1} (W_{x_1 x_0 |y_0}+W_{y_1 y_0 |x_0}),\nonumber\\
q(x_1,y_1):=&W_{y_1 y_0 |x_1}W_{x_1 x_0 |y_0}  (W_{x_1 x_0 |y_1}+W_{y_0 y_1 |x_0})\nonumber\\
&+W_{x_1 x_0 |y_1} W_{y_1 y_0 |x_0} (W_{x_0 x_1 |y_0}+W_{y_1 y_0 |x_1}).
\end{align}

If $\Gamma := \Gamma^{(\nu)}_{y_i}=\Gamma^{(\nu)}_{x_i}$ does not depend on $\nu$, $x_i$, $y_i$, we can simplify the steady solution as
\begin{align}
p_{\rm{ss}}(x_0, y_0)&=N(2-(f^{(XL)}_X+f^{(XR)}_X))(2-(f^{(YL)}_Y+f^{(YR)}_Y)),\nonumber\\
p_{\rm{ss}}(x_1, y_0)&=N(f^{(XL)}_X+f^{(XR)}_X)(2-(f^{(YL)}_Y+f^{(YR)}_Y)),\nonumber\\
p_{\rm{ss}}(x_0, y_1)&=N(2-(f^{(XL)}_X+f^{(XR)}_X))(f^{(YL)}_Y+f^{(YR)}_Y),\nonumber\\
p_{\rm{ss}}(x_1, y_1)&=N(f^{(XL)}_X+f^{(XR)}_X)(f^{(YL)}_Y+f^{(YR)}_Y),\nonumber\\
\end{align}
where $N>0$ is a normalization constant.
In this special case,  we have
\begin{align}
\frac{p_{\rm ss}(x_0, y_1)p_{\rm ss}(x_1, y_0)}{p_{\rm ss}(x_0, y_0)p_{\rm ss}(x_1, y_1)} = 1,
\end{align}
and therefore 
\begin{equation}
F_I=0
\end{equation}
for any $F_X$ and $F_Y$.


\begin{thebibliography}{66}%
\makeatletter
\providecommand \@ifxundefined [1]{%
 \@ifx{#1\undefined}
}%
\providecommand \@ifnum [1]{%
 \ifnum #1\expandafter \@firstoftwo
 \else \expandafter \@secondoftwo
 \fi
}%
\providecommand \@ifx [1]{%
 \ifx #1\expandafter \@firstoftwo
 \else \expandafter \@secondoftwo
 \fi
}%
\providecommand \natexlab [1]{#1}%
\providecommand \enquote  [1]{``#1''}%
\providecommand \bibnamefont  [1]{#1}%
\providecommand \bibfnamefont [1]{#1}%
\providecommand \citenamefont [1]{#1}%
\providecommand \href@noop [0]{\@secondoftwo}%
\providecommand \href [0]{\begingroup \@sanitize@url \@href}%
\providecommand \@href[1]{\@@startlink{#1}\@@href}%
\providecommand \@@href[1]{\endgroup#1\@@endlink}%
\providecommand \@sanitize@url [0]{\catcode `\\12\catcode `\$12\catcode
  `\&12\catcode `\#12\catcode `\^12\catcode `\_12\catcode `\%12\relax}%
\providecommand \@@startlink[1]{}%
\providecommand \@@endlink[0]{}%
\providecommand \url  [0]{\begingroup\@sanitize@url \@url }%
\providecommand \@url [1]{\endgroup\@href {#1}{\urlprefix }}%
\providecommand \urlprefix  [0]{URL }%
\providecommand \Eprint [0]{\href }%
\providecommand \doibase [0]{http://dx.doi.org/}%
\providecommand \selectlanguage [0]{\@gobble}%
\providecommand \bibinfo  [0]{\@secondoftwo}%
\providecommand \bibfield  [0]{\@secondoftwo}%
\providecommand \translation [1]{[#1]}%
\providecommand \BibitemOpen [0]{}%
\providecommand \bibitemStop [0]{}%
\providecommand \bibitemNoStop [0]{.\EOS\space}%
\providecommand \EOS [0]{\spacefactor3000\relax}%
\providecommand \BibitemShut  [1]{\csname bibitem#1\endcsname}%
\let\auto@bib@innerbib\@empty
%</preamble>
\bibitem [{\citenamefont {Onsager}(1931{\natexlab{a}})}]{Onsager1}%
  \BibitemOpen
  \bibfield  {author} {\bibinfo {author} {\bibfnamefont {L.}~\bibnamefont
  {Onsager}},\ } {\bibfield
  {journal} {\bibinfo  {journal} {Phys. Rev.}\ }\textbf {\bibinfo {volume}
  {37}},\ \bibinfo {pages} {405} (\bibinfo {year}
  {1931}{\natexlab{a}})}\BibitemShut {NoStop}%
\bibitem [{\citenamefont {Onsager}(1931{\natexlab{b}})}]{Onsager2}%
  \BibitemOpen
  \bibfield  {author} {\bibinfo {author} {\bibfnamefont {L.}~\bibnamefont
  {Onsager}},\ } {\bibfield
  {journal} {\bibinfo  {journal} {Phys. Rev.}\ }\textbf {\bibinfo {volume}
  {38}},\ \bibinfo {pages} {2265} (\bibinfo {year}
  {1931}{\natexlab{b}})}\BibitemShut {NoStop}%
\bibitem [{\citenamefont {Callen}(1948)}]{Callen}%
  \BibitemOpen
  \bibfield  {author} {\bibinfo {author} {\bibfnamefont {H.~B.}\ \bibnamefont
  {Callen}},\ } {\bibfield  {journal} {\bibinfo
  {journal} {Phys. Rev.}\ }\textbf {\bibinfo {volume} {73}},\ \bibinfo {pages}
  {1349} (\bibinfo {year} {1948})}\BibitemShut {NoStop}%
\bibitem [{\citenamefont {Goupil}\ \emph {et~al.}(2011)\citenamefont {Goupil},
  \citenamefont {Seifert}, \citenamefont {Zabrocki},\ and\ \citenamefont
  {M{\"u}ller}}]{GoupilSnyder}%
  \BibitemOpen
  \bibfield  {author} {\bibinfo {author} {\bibfnamefont {C.}~\bibnamefont
  {Goupil}}, \bibinfo {author} {\bibfnamefont {W}~\bibnamefont {Seifert}},
  \bibinfo {author} {\bibfnamefont {K.}~\bibnamefont {Zabrocki}},
  \bibinfo {author} {\bibfnamefont {E.}\ \bibnamefont {M{\"u}ller}, \ and\
  \bibfnamefont {~G.~J. Snyder}},\ } {\bibfield  {journal} {\bibinfo  {journal} {Entropy}\ }\textbf
  {\bibinfo {volume} {13}},\ \bibinfo {pages} {1481} (\bibinfo {year}
  {2011})}\BibitemShut {NoStop}%
\bibitem [{\citenamefont {Brandner}\ \emph {et~al.}(2013)\citenamefont
  {Brandner}, \citenamefont {Saito},\ and\ \citenamefont
  {Seifert}}]{BrandnerSeifert}%
  \BibitemOpen
  \bibfield  {author} {\bibinfo {author} {\bibfnamefont {K.}~\bibnamefont
  {Brandner}}, \bibinfo {author} {\bibfnamefont {K.}~\bibnamefont {Saito}}, \
  and\ \bibinfo {author} {\bibfnamefont {U.}~\bibnamefont {Seifert}},\
  } {\bibfield  {journal} {\bibinfo  {journal}
  {Phys. Rev. Lett.}\ }\textbf {\bibinfo {volume} {110}},\ \bibinfo {pages}
  {070603} (\bibinfo {year} {2013})}\BibitemShut {NoStop}%
\bibitem [{\citenamefont {Kedem}\ and\ \citenamefont
  {Katchalsky}(1958)}]{KedemKatchalsky}%
  \BibitemOpen
  \bibfield  {author} {\bibinfo {author} {\bibfnamefont {O.}~\bibnamefont
  {Kedem}}\ and\ \bibinfo {author} {\bibfnamefont {A.}~\bibnamefont
  {Katchalsky}},\ } {\bibfield  {journal} {\bibinfo
  {journal} {Biochim. Biophys.}\ }\textbf {\bibinfo {volume} {27}},\ \bibinfo
  {pages} {229} (\bibinfo {year} {1958})}\BibitemShut {NoStop}%
\bibitem [{\citenamefont {Mason}\ and\ \citenamefont
  {Lonsdale}(1990)}]{MasonLonsdale}%
  \BibitemOpen
  \bibfield  {author} {\bibinfo {author} {\bibfnamefont {E.~A.}\ \bibnamefont
  {Mason}}\ and\ \bibinfo {author} {\bibfnamefont {H.~K.}\ \bibnamefont
  {Lonsdale}},\ }
  {\bibfield  {journal} {\bibinfo  {journal} {J. Membr. Sci.}\ }\textbf
  {\bibinfo {volume} {51}},\ \bibinfo {pages} {1} (\bibinfo {year}
  {1990})}\BibitemShut {NoStop}%
\bibitem [{\citenamefont {Baranowski}(1991)}]{Baranowski}%
  \BibitemOpen
  \bibfield  {author} {\bibinfo {author} {\bibfnamefont {B.}~\bibnamefont
  {Baranowski}},\ } {\bibfield  {journal} {\bibinfo  {journal} {Journal of
  Membrane Science}\ }\textbf {\bibinfo {volume} {57}},\ \bibinfo {pages}
  {119} (\bibinfo {year} {1991})}\BibitemShut {NoStop}%
\bibitem [{\citenamefont {Chu}\ \emph {et~al.}(1985)\citenamefont {Chu},
  \citenamefont {Gisser}, \citenamefont {Kupferschmid},\ and\ \citenamefont
  {Zelman}}]{Richard}%
  \BibitemOpen
  \bibfield  {author} {\bibinfo {author} {\bibfnamefont {R.}~\bibnamefont
  {Chu}}, \bibinfo {author} {\bibfnamefont {D.}~\bibnamefont {Gisser}},
  \bibinfo {author} {\bibfnamefont {M.}~\bibnamefont {Kupferschmid}}, \ and\
  \bibinfo {author} {\bibfnamefont {A.}~\bibnamefont {Zelman}},\ } {\bibfield  {journal} {\bibinfo  {journal} {Journal of
  Membrane Science}\ }\textbf {\bibinfo {volume} {22}},\ \bibinfo {pages}
  {77} (\bibinfo {year} {1985})}\BibitemShut {NoStop}%
\bibitem [{\citenamefont {Gerber}\ \emph {et~al.}(2016)\citenamefont {Gerber},
  \citenamefont {Fr\"{o}hlich}, \citenamefont {Lichtenberg-Frat\'{e}},
  \citenamefont {Shabala}, \citenamefont {Shabala},\ and\ \citenamefont
  {Klipp}}]{Gerber}%
  \BibitemOpen
  \bibfield  {author} {\bibinfo {author} {\bibfnamefont {S.}~\bibnamefont
  {Gerber}}, \bibinfo {author} {\bibfnamefont {M.}~\bibnamefont
  {Fr\"{o}hlich}}, \bibinfo {author} {\bibfnamefont {H.}~\bibnamefont
  {Lichtenberg-Frat\'{e}}}, \bibinfo {author} {\bibfnamefont {S.}~\bibnamefont
  {Shabala}}, \bibinfo {author} {\bibfnamefont {L.}~\bibnamefont {Shabala}}, \
  and\ \bibinfo {author} {\bibfnamefont {E.}~\bibnamefont {Klipp}},\ }
  {\bibfield  {journal} {\bibinfo  {journal} {PLoS Comput. Biol.}\ }\textbf
  {\bibinfo {volume} {12}},\ \bibinfo {pages} {e1004703} (\bibinfo {year}
  {2016})}\BibitemShut {NoStop}%
\bibitem [{\citenamefont {Schnakenberg}(1976)}]{Schnakenberg}%
  \BibitemOpen
  \bibfield  {author} {\bibinfo {author} {\bibfnamefont {J.}~\bibnamefont
  {Schnakenberg}},\ } {\bibfield  {journal} {\bibinfo  {journal} {Rev. Mod. Phys.}\
  }\textbf {\bibinfo {volume} {48}},\ \bibinfo {pages} {571} (\bibinfo {year}
  {1976})}\BibitemShut {NoStop}%
\bibitem [{\citenamefont {Sekimoto}(2010)}]{Sekimoto}%
  \BibitemOpen
  \bibfield  {author} {\bibinfo {author} {\bibfnamefont {K.}~\bibnamefont
  {Sekimoto}},\ }\href@noop {} {\emph {\bibinfo {title} {Stochastic
 Energetics}}}, Lecture Notes in Physics,\ vol.\ \bibinfo {volume} {799}\ (\bibinfo  {publisher}
  {Springer, New York,},\ \bibinfo {year} {2010})\BibitemShut {NoStop}%
\bibitem [{\citenamefont {Seifert}(2012)}]{Seifert}%
  \BibitemOpen
  \bibfield  {author} {\bibinfo {author} {\bibfnamefont {U.}~\bibnamefont
  {Seifert}},\ } {\bibfield  {journal} {\bibinfo  {journal} {Rep. Prog. Phys.}\ }\textbf
  {\bibinfo {volume} {75}},\ \bibinfo {pages} {126001} (\bibinfo {year}
  {2012})}\BibitemShut {NoStop}%
\bibitem [{\citenamefont {Andrieux}\ and\ \citenamefont
  {Gaspard}(2004)}]{Andrieux}%
  \BibitemOpen
  \bibfield  {author} {\bibinfo {author} {\bibfnamefont {D.}~\bibnamefont
  {Andrieux}}\ and\ \bibinfo {author} {\bibfnamefont {P.}~\bibnamefont
  {Gaspard}},\ } {\bibfield
  {journal} {\bibinfo  {journal} {J. Chem. Phys.}\ }\textbf {\bibinfo {volume}
  {121}},\ \bibinfo {pages} {6167} (\bibinfo {year} {2004})}\BibitemShut
  {NoStop}%
\bibitem [{\citenamefont {Andrieux}\ and\ \citenamefont
  {Gaspard}(2007{\natexlab{a}})}]{Andriuex2}%
  \BibitemOpen
  \bibfield  {author} {\bibinfo {author} {\bibfnamefont {D.}~\bibnamefont
  {Andrieux}}\ and\ \bibinfo {author} {\bibfnamefont {P.}~\bibnamefont
  {Gaspard}},\ } 
  {\bibfield  {journal} {\bibinfo  {journal} {J. Stat. Phys.}\ }\textbf
  {\bibinfo {volume} {127}},\ \bibinfo {pages} {107} (\bibinfo {year}
  {2007}{\natexlab{a}})}\BibitemShut {NoStop}%
\bibitem [{\citenamefont {Andrieux}\ and\ \citenamefont
  {Gaspard}(2007{\natexlab{b}})}]{Andrieux3}%
  \BibitemOpen
  \bibfield  {author} {\bibinfo {author} {\bibfnamefont {D.}~\bibnamefont
  {Andrieux}}\ and\ \bibinfo {author} {\bibfnamefont {P.}~\bibnamefont
  {Gaspard}},\ } 
  {\bibfield  {journal} {\bibinfo  {journal} {J. Stat. Mech.: Theor. Exp.}\
  }\textbf {\bibinfo {volume} {2007}},\ \bibinfo {pages} {P02006} (\bibinfo
  {year} {2007}{\natexlab{b}})}\BibitemShut {NoStop}%
\bibitem [{\citenamefont {Saito}\ and\ \citenamefont
  {Utsumi}(2008)}]{SaitoUtsumi}%
  \BibitemOpen
  \bibfield  {author} {\bibinfo {author} {\bibfnamefont {K.}~\bibnamefont
  {Saito}}\ and\ \bibinfo {author} {\bibfnamefont {Y.}~\bibnamefont {Utsumi}},\
  } 
  {\bibfield  {journal} {\bibinfo  {journal} {Phys. Rev. B}\ }\textbf {\bibinfo
  {volume} {78}},\ \bibinfo {pages} {115429} (\bibinfo {year}
  {2008})}\BibitemShut {NoStop}%
\bibitem [{\citenamefont {Curzon}\ and\ \citenamefont
  {Ahlborn}(1975)}]{CurzonAhlborn}%
  \BibitemOpen
  \bibfield  {author} {\bibinfo {author} {\bibfnamefont {F.~L.}\ \bibnamefont
  {Curzon}}\ and\ \bibinfo {author} {\bibfnamefont {B.}~\bibnamefont
  {Ahlborn}},\ } {\bibfield
  {journal} {\bibinfo  {journal} {Am. J. Phys.}\ }\textbf {\bibinfo {volume}
  {43}},\ \bibinfo {pages} {22} (\bibinfo {year} {1975})}\BibitemShut
  {NoStop}%
\bibitem [{\citenamefont {Van~den Broeck}(2005)}]{Broeck}%
  \BibitemOpen
  \bibfield  {author} {\bibinfo {author} {\bibfnamefont {C.}~\bibnamefont
  {Van~den Broeck}},\ }  {\bibfield
  {journal} {\bibinfo  {journal} {Phys. Rev. Lett.}\ }\textbf {\bibinfo
  {volume} {95}},\ \bibinfo {pages} {190602} (\bibinfo {year}
  {2005})}\BibitemShut {NoStop}%
\bibitem [{\citenamefont {Benenti}\ \emph {et~al.}(2011)\citenamefont
  {Benenti}, \citenamefont {Saito},\ and\ \citenamefont
  {Casati}}]{BenentiCasati}%
  \BibitemOpen
  \bibfield  {author} {\bibinfo {author} {\bibfnamefont {G.}~\bibnamefont
  {Benenti}}, \bibinfo {author} {\bibfnamefont {K.}~\bibnamefont {Saito}}, \
  and\ \bibinfo {author} {\bibfnamefont {G.}~\bibnamefont {Casati}},\
  }
  {\bibfield  {journal} {\bibinfo  {journal} {Phys. Rev. Lett.}\ }\textbf
  {\bibinfo {volume} {106}},\ \bibinfo {pages} {230602} (\bibinfo {year}
  {2011})}\BibitemShut {NoStop}%
\bibitem [{\citenamefont {Proesmans}\ and\ \citenamefont {Van~den
  Broeck}(2015)}]{ProesmansBroeck}%
  \BibitemOpen
  \bibfield  {author} {\bibinfo {author} {\bibfnamefont {K.}~\bibnamefont
  {Proesmans}}\ and\ \bibinfo {author} {\bibfnamefont {C.}~\bibnamefont
  {Van~den Broeck}},\ } {\bibfield
  {journal} {\bibinfo  {journal} {Phys. Rev. Lett.}\ }\textbf {\bibinfo
  {volume} {115}},\ \bibinfo {pages} {090601} (\bibinfo {year}
  {2015})}\BibitemShut {NoStop}%
\bibitem [{\citenamefont {Brandner}\ \emph {et~al.}(2015)\citenamefont
  {Brandner}, \citenamefont {Saito},\ and\ \citenamefont
  {Seifert}}]{BrandnerSeifert2}%
  \BibitemOpen
  \bibfield  {author} {\bibinfo {author} {\bibfnamefont {K.}~\bibnamefont
  {Brandner}}, \bibinfo {author} {\bibfnamefont {K.}~\bibnamefont {Saito}}, \
  and\ \bibinfo {author} {\bibfnamefont {U.}~\bibnamefont {Seifert}},\
  }
  {\bibfield  {journal} {\bibinfo  {journal} {Phys. Rev. X}\ }\textbf {\bibinfo
  {volume} {5}},\ \bibinfo {pages} {031019} (\bibinfo {year}
  {2015})}\BibitemShut {NoStop}%
\bibitem [{\citenamefont {Parrondo}\ \emph {et~al.}(2015)\citenamefont
  {Parrondo}, \citenamefont {Horowitz},\ and\ \citenamefont {Sagawa}}]{review}%
  \BibitemOpen
  \bibfield  {author} {\bibinfo {author} {\bibfnamefont {J.~M.~R.}\
  \bibnamefont {Parrondo}}, \bibinfo {author} {\bibfnamefont {J.~M.}\
  \bibnamefont {Horowitz}}, \ and\ \bibinfo {author} {\bibfnamefont
  {T.}~\bibnamefont {Sagawa}},\ } {\bibfield  {journal}
  {\bibinfo  {journal} {Nat. Phys.}\ }\textbf {\bibinfo {volume} {11}},\
  \bibinfo {pages} {131} (\bibinfo {year} {2015})}\BibitemShut {NoStop}%
\bibitem [{\citenamefont {Touchette}\ and\ \citenamefont
  {Lloyd}(2000)}]{TouchetteLloyd}%
  \BibitemOpen
  \bibfield  {author} {\bibinfo {author} {\bibfnamefont {H.}~\bibnamefont
  {Touchette}}\ and\ \bibinfo {author} {\bibfnamefont {S.}~\bibnamefont
  {Lloyd}},\ } {\bibfield
  {journal} {\bibinfo  {journal} {Phys. Rev. Lett.}\ }\textbf {\bibinfo
  {volume} {84}},\ \bibinfo {pages} {1156} (\bibinfo {year}
  {2000})}\BibitemShut {NoStop}%
\bibitem [{\citenamefont {Sagawa}\ and\ \citenamefont
  {Ueda}(2008)}]{SagawaUeda08}%
  \BibitemOpen
  \bibfield  {author} {\bibinfo {author} {\bibfnamefont {T.}~\bibnamefont
  {Sagawa}}\ and\ \bibinfo {author} {\bibfnamefont {M.}~\bibnamefont {Ueda}},\
  } {\bibfield
  {journal} {\bibinfo  {journal} {Phys. Rev. Lett.}\ }\textbf {\bibinfo
  {volume} {100}},\ \bibinfo {pages} {080403} (\bibinfo {year}
  {2008})}\BibitemShut {NoStop}%
\bibitem [{\citenamefont {Sagawa}\ and\ \citenamefont
  {Ueda}(2011)}]{SagawaUeda09}%
  \BibitemOpen
  \bibfield  {author} {\bibinfo {author} {\bibfnamefont {T.}~\bibnamefont
  {Sagawa}}\ and\ \bibinfo {author} {\bibfnamefont {M.}~\bibnamefont {Ueda}},\
  } {\bibfield
  {journal} {\bibinfo  {journal} {Phys. Rev. Lett.}\ }\textbf {\bibinfo
  {volume} {106}},\ \bibinfo {pages} {189901} \ (E) (\bibinfo {year}
  {2011})}\BibitemShut {NoStop}%
\bibitem [{\citenamefont {Allahverdyan}\ \emph {et~al.}(2009)\citenamefont
  {Allahverdyan}, \citenamefont {Janzing},\ and\ \citenamefont
  {Mahler}}]{AllahverdyanMahler}%
  \BibitemOpen
  \bibfield  {author} {\bibinfo {author} {\bibfnamefont {A.~E.}\ \bibnamefont
  {Allahverdyan}}, \bibinfo {author} {\bibfnamefont {D.}~\bibnamefont
  {Janzing}}, \ and\ \bibinfo {author} {\bibfnamefont {G.}~\bibnamefont
  {Mahler}},\ } {\bibfield
  {journal} {\bibinfo  {journal} {J. Stat. Mech.: Theor. Exp.}\ }\textbf
  {\bibinfo {volume} {2009}},\ \bibinfo {pages} {P09011} (\bibinfo {year}
  {2009})}\BibitemShut {NoStop}%
\bibitem [{\citenamefont {Sagawa}\ and\ \citenamefont
  {Ueda}(2010)}]{SagawaUeda2}%
  \BibitemOpen
  \bibfield  {author} {\bibinfo {author} {\bibfnamefont {T.}~\bibnamefont
  {Sagawa}}\ and\ \bibinfo {author} {\bibfnamefont {M.}~\bibnamefont {Ueda}},\
  } {\bibfield
  {journal} {\bibinfo  {journal} {Phys. Rev. Lett.}\ }\textbf {\bibinfo
  {volume} {104}},\ \bibinfo {pages} {090602} (\bibinfo {year}
  {2010})}\BibitemShut {NoStop}%
\bibitem [{\citenamefont {Fujitani}\ and\ \citenamefont
  {Suzuki}(2010)}]{FujitaniSuzuki}%
  \BibitemOpen
  \bibfield  {author} {\bibinfo {author} {\bibfnamefont {Y.}~\bibnamefont
  {Fujitani}}\ and\ \bibinfo {author} {\bibfnamefont {H.}~\bibnamefont
  {Suzuki}},\ } {\bibfield
   {journal} {\bibinfo  {journal} {J. Phys. Soc. Japan}\ }\textbf {\bibinfo
  {volume} {79}},\ \bibinfo {pages} {104003} (\bibinfo {year}
  {2010})}\BibitemShut {NoStop}%
\bibitem [{\citenamefont {Horowitz}\ and\ \citenamefont
  {Vaikuntanathan}(2010)}]{HorowitzVaikuntanathan}%
  \BibitemOpen
  \bibfield  {author} {\bibinfo {author} {\bibfnamefont {J.~M.}\ \bibnamefont
  {Horowitz}}\ and\ \bibinfo {author} {\bibfnamefont {S.}~\bibnamefont
  {Vaikuntanathan}},\ } {\bibfield  {journal} {\bibinfo  {journal} {Phys.
  Rev. E}\ }\textbf {\bibinfo {volume} {82}},\ \bibinfo {pages} {061120}
  (\bibinfo {year} {2010})}\BibitemShut {NoStop}%
\bibitem [{\citenamefont {Sagawa}\ and\ \citenamefont
  {Ueda}(2012)}]{SagawaUeda4}%
  \BibitemOpen
  \bibfield  {author} {\bibinfo {author} {\bibfnamefont {T.}~\bibnamefont
  {Sagawa}}\ and\ \bibinfo {author} {\bibfnamefont {M.}~\bibnamefont {Ueda}},\
  }{\bibfield  {journal} {\bibinfo  {journal} {Phys. Rev. Lett.}\
  }\textbf {\bibinfo {volume} {109}},\ \bibinfo {pages} {180602} (\bibinfo
  {year} {2012})}\BibitemShut {NoStop}%
\bibitem [{\citenamefont {Toyabe}\ \emph {et~al.}(2010)\citenamefont {Toyabe},
  \citenamefont {Ueda}, \citenamefont {Muneyuki},\ and\ \citenamefont
  {Sano}}]{Toyabe}%
  \BibitemOpen
  \bibfield  {author} {\bibinfo {author} {\bibfnamefont {S.}~\bibnamefont
  {Toyabe}, \bibfnamefont {T.~Sagawa}}, \bibinfo {author} {\bibfnamefont
  {M.}~\bibnamefont {Ueda}}, \bibinfo {author} {\bibfnamefont {E.}~\bibnamefont
  {Muneyuki}}, \ and\ \bibinfo {author} {\bibfnamefont {M.}~\bibnamefont
  {Sano}},\ } {\bibfield  {journal}
  {\bibinfo  {journal} {Nat. Phys.}\ }\textbf {\bibinfo {volume} {6}},\
  \bibinfo {pages} {988} (\bibinfo {year} {2010})}\BibitemShut {NoStop}%
\bibitem [{\citenamefont {B{\'e}rut}\ \emph {et~al.}(2012)\citenamefont
  {B{\'e}rut}, \citenamefont {Arakelyan}, \citenamefont {Petrosyan},
  \citenamefont {Ciliberto}, \citenamefont {Dillenschneider},\ and\
  \citenamefont {Lutz}}]{Berut}%
  \BibitemOpen
  \bibfield  {author} {\bibinfo {author} {\bibfnamefont {A.}~\bibnamefont
  {B{\'e}rut}}, \bibinfo {author} {\bibfnamefont {A.}~\bibnamefont
  {Arakelyan}}, \bibinfo {author} {\bibfnamefont {A.}~\bibnamefont
  {Petrosyan}}, \bibinfo {author} {\bibfnamefont {S.}~\bibnamefont
  {Ciliberto}}, \bibinfo {author} {\bibfnamefont {R.}~\bibnamefont
  {Dillenschneider}}, \ and\ \bibinfo {author} {\bibfnamefont {E.}~\bibnamefont
  {Lutz}},\ }{\bibfield  {journal} {\bibinfo  {journal}
  {Nature\ (London)}\ }\textbf {\bibinfo {volume} {483}},\ \bibinfo {pages} {187}
  (\bibinfo {year} {2012})}\BibitemShut {NoStop}%
\bibitem [{\citenamefont {Koski}\ \emph {et~al.}(2014)\citenamefont {Koski},
  \citenamefont {Maisi}, \citenamefont {Sagawa},\ and\ \citenamefont
  {Pekola}}]{Koski}%
  \BibitemOpen
  \bibfield  {author} {\bibinfo {author} {\bibfnamefont {J.~V}\ \bibnamefont
  {Koski}}, \bibinfo {author} {\bibfnamefont {V.~F.}\ \bibnamefont {Maisi}},
  \bibinfo {author} {\bibfnamefont {T.}~\bibnamefont {Sagawa}}, \ and\ \bibinfo
  {author} {\bibfnamefont {J.~P.}\ \bibnamefont {Pekola}},\ }{\bibfield  {journal} {\bibinfo  {journal} {Phys. Rev. Lett.}\ }\textbf
  {\bibinfo {volume} {113}},\ \bibinfo {pages} {030601} (\bibinfo {year}
  {2014})}\BibitemShut {NoStop}%
\bibitem [{\citenamefont {Mehta}\ \emph {et~al.}(2016)\citenamefont {Mehta},
  \citenamefont {Lang},\ and\ \citenamefont {Schwab}}]{MehtaSchwab2}%
  \BibitemOpen
  \bibfield  {author} {\bibinfo {author} {\bibfnamefont {P.}~\bibnamefont
  {Mehta}}, \bibinfo {author} {\bibfnamefont {A.~H.}\ \bibnamefont {Lang}}, \
  and\ \bibinfo {author} {\bibfnamefont {D.~J.}\ \bibnamefont {Schwab}},\
  }  {\bibfield  {journal} {\bibinfo
  {journal} {J. Stat. Phys.}\ }\textbf {\bibinfo {volume} {162}},\ \bibinfo
  {pages} {1153} (\bibinfo {year} {2016})}\BibitemShut {NoStop}%
\bibitem [{\citenamefont {Ito}\ and\ \citenamefont
  {Sagawa}(2015)}]{ItoSagawa2}%
  \BibitemOpen
  \bibfield  {author} {\bibinfo {author} {\bibfnamefont {S.}~\bibnamefont
  {Ito}}\ and\ \bibinfo {author} {\bibfnamefont {T.}~\bibnamefont {Sagawa}},\
  }
  {\bibfield  {journal} {\bibinfo  {journal} {Nat. Commun.}}\ \textbf {\bibinfo
  {volume} {6}}, {\bibinfo
  {pages} {7498}} (\bibinfo {year} {2015})}\BibitemShut {NoStop}%
\bibitem [{\citenamefont {Barato}\ \emph {et~al.}(2014)\citenamefont {Barato},
  \citenamefont {Hartich},\ and\ \citenamefont {Seifert}}]{HartichSeifert}%
  \BibitemOpen
  \bibfield  {author} {\bibinfo {author} {\bibfnamefont {A.~C.}\ \bibnamefont
  {Barato}}, \bibinfo {author} {\bibfnamefont {D.}~\bibnamefont {Hartich}}, \
  and\ \bibinfo {author} {\bibfnamefont {U.}~\bibnamefont {Seifert}},\
  } {\bibfield  {journal} {\bibinfo
  {journal} {New. J. Phys.}\ }\textbf {\bibinfo {volume} {16}},\ \bibinfo
  {pages} {103024} (\bibinfo {year} {2014})}\BibitemShut {NoStop}%
\bibitem [{\citenamefont {Sartori}\ \emph {et~al.}(2014)\citenamefont
  {Sartori}, \citenamefont {Granger}, \citenamefont {Lee},\ and\ \citenamefont
  {Horowitz}}]{SartoriHorowitz}%
  \BibitemOpen
  \bibfield  {author} {\bibinfo {author} {\bibfnamefont {P.}~\bibnamefont
  {Sartori}}, \bibinfo {author} {\bibfnamefont {L.}~\bibnamefont {Granger}},
  \bibinfo {author} {\bibfnamefont {C.~F.}\ \bibnamefont {Lee}}, \ and\
  \bibinfo {author} {\bibfnamefont {J.~M.}\ \bibnamefont {Horowitz}},\
  } {\bibfield
  {journal} {\bibinfo  {journal} {PLoS Comput. Biol.}\ }\textbf {\bibinfo
  {volume} {10}},\ \bibinfo {pages} {e1003974} (\bibinfo {year}
  {2014})}\BibitemShut {NoStop}%
\bibitem [{\citenamefont {Ouldridge}\ \emph {et~al.}(2015)\citenamefont
  {Ouldridge}, \citenamefont {Govern},\ and\ \citenamefont {ten
  Wolde}}]{tenWolde}%
  \BibitemOpen
  \bibfield  {author} {\bibinfo {author} {\bibfnamefont {T.~E.}\ \bibnamefont
  {Ouldridge}}, \bibinfo {author} {\bibfnamefont {C.~C.}\ \bibnamefont
  {Govern}}, \ and\ \bibinfo {author} {\bibfnamefont {P.~R.}\ \bibnamefont {ten
  Wolde}},\ } 
  {\bibfield  {journal} {\bibinfo  {journal} {arXiv preprint arXiv:1503.00909}\
  } (\bibinfo {year} {2015})}\BibitemShut {NoStop}%
\bibitem [{\citenamefont {Lan}\ \emph {et~al.}(2012)\citenamefont {Lan},
  \citenamefont {Sartori}, \citenamefont {Neumann}, \citenamefont {Sourjik},\
  and\ \citenamefont {Tu}}]{Lan}%
  \BibitemOpen
  \bibfield  {author} {\bibinfo {author} {\bibfnamefont {G.}~\bibnamefont
  {Lan}}, \bibinfo {author} {\bibfnamefont {P.}~\bibnamefont {Sartori}},
  \bibinfo {author} {\bibfnamefont {S.}~\bibnamefont {Neumann}}, \bibinfo
  {author} {\bibfnamefont {V.}~\bibnamefont {Sourjik}}, \ and\ \bibinfo
  {author} {\bibfnamefont {Y.}~\bibnamefont {Tu}},\ } {\bibfield  {journal} {\bibinfo  {journal}
  {Nat. Phys.}\ }\textbf {\bibinfo {volume} {8}},\ \bibinfo {pages} {422}
  (\bibinfo {year} {2012})}\BibitemShut {NoStop}%
\bibitem [{\citenamefont {Mehta}\ and\ \citenamefont {Schwab}(2012)}]{Mehta}%
  \BibitemOpen
  \bibfield  {author} {\bibinfo {author} {\bibfnamefont {P.}~\bibnamefont
  {Mehta}}\ and\ \bibinfo {author} {\bibfnamefont {D.~J.}\ \bibnamefont
  {Schwab}},\ } {\bibfield  {journal} {\bibinfo
  {journal} {Proc. Natl. Acad. Sci. U.S.A.}\ }\textbf {\bibinfo {volume}
  {109}},\ \bibinfo {pages} {17978} (\bibinfo {year}
  {2012})}\BibitemShut {NoStop}%
\bibitem [{\citenamefont {Lang}\ \emph {et~al.}(2014)\citenamefont {Lang},
  \citenamefont {Fisher}, \citenamefont {Mora},\ and\ \citenamefont
  {Mehta}}]{LangMehta}%
  \BibitemOpen
  \bibfield  {author} {\bibinfo {author} {\bibfnamefont {A.~H.}\ \bibnamefont
  {Lang}}, \bibinfo {author} {\bibfnamefont {C.~K.}\ \bibnamefont {Fisher}},
  \bibinfo {author} {\bibfnamefont {T.}~\bibnamefont {Mora}}, \ and\ \bibinfo
  {author} {\bibfnamefont {P.}~\bibnamefont {Mehta}},\ }  {\bibfield  {journal} {\bibinfo  {journal} {Phys.
  Rev. Lett.}\ }\textbf {\bibinfo {volume} {113}},\ \bibinfo {pages} {148103}
  (\bibinfo {year} {2014})}\BibitemShut {NoStop}%
\bibitem [{\citenamefont {Mandal}\ and\ \citenamefont
  {Jarzynski}(2012)}]{MandalJarzynski}%
  \BibitemOpen
  \bibfield  {author} {\bibinfo {author} {\bibfnamefont {D.}~\bibnamefont
  {Mandal}}\ and\ \bibinfo {author} {\bibfnamefont {C.}~\bibnamefont
  {Jarzynski}},\ }  {\bibfield  {journal} {\bibinfo  {journal} {Proc. Natl. Acad. Sci.
  U.S.A.}\ }\textbf {\bibinfo {volume} {109}},\ \bibinfo {pages} {11641}
  (\bibinfo {year} {2012})}\BibitemShut {NoStop}%
\bibitem [{\citenamefont {Strasberg}\ \emph {et~al.}(2013)\citenamefont
  {Strasberg}, \citenamefont {Schaller}, \citenamefont {Brandes},\ and\
  \citenamefont {Esposito}}]{StrasbergEsposito}%
  \BibitemOpen
  \bibfield  {author} {\bibinfo {author} {\bibfnamefont {P.}~\bibnamefont
  {Strasberg}}, \bibinfo {author} {\bibfnamefont {G.}~\bibnamefont {Schaller}},
  \bibinfo {author} {\bibfnamefont {T.}~\bibnamefont {Brandes}}, \ and\
  \bibinfo {author} {\bibfnamefont {M.}~\bibnamefont {Esposito}},\ } {\bibfield  {journal}
  {\bibinfo  {journal} {Phys. Rev. Lett.}\ }\textbf {\bibinfo {volume} {110}},\
  \bibinfo {pages} {040601} (\bibinfo {year} {2013})}\BibitemShut {NoStop}%
\bibitem [{\citenamefont {Horowitz}\ \emph {et~al.}(2013)\citenamefont
  {Horowitz}, \citenamefont {Sagawa},\ and\ \citenamefont
  {Parrondo}}]{HorowitzParrondo2}%
  \BibitemOpen
  \bibfield  {author} {\bibinfo {author} {\bibfnamefont {J.~M.}\ \bibnamefont
  {Horowitz}}, \bibinfo {author} {\bibfnamefont {T.}~\bibnamefont {Sagawa}}, \
  and\ \bibinfo {author} {\bibfnamefont {J.~M.~R.}\ \bibnamefont {Parrondo}},\
  } {\bibfield  {journal}
  {\bibinfo  {journal} {Phys. Rev. Lett.}\ }\textbf {\bibinfo {volume} {111}},\
  \bibinfo {pages} {010602} (\bibinfo {year} {2013})}\BibitemShut {NoStop}%
\bibitem [{\citenamefont {Deffner}\ and\ \citenamefont
  {Jarzynski}(2013)}]{DeffnerJarzynski}%
  \BibitemOpen
  \bibfield  {author} {\bibinfo {author} {\bibfnamefont {S.}~\bibnamefont
  {Deffner}}\ and\ \bibinfo {author} {\bibfnamefont {C.}~\bibnamefont
  {Jarzynski}},\ }{\bibfield  {journal} {\bibinfo  {journal} {Phys.
  Rev. X}\ }\textbf {\bibinfo {volume} {3}},\ \bibinfo {pages} {041003}
  (\bibinfo {year} {2013})}\BibitemShut {NoStop}%
\bibitem [{\citenamefont {Ito}\ and\ \citenamefont {Sagawa}(2013)}]{ItoSagawa}%
  \BibitemOpen
  \bibfield  {author} {\bibinfo {author} {\bibfnamefont {S.}~\bibnamefont
  {Ito}}\ and\ \bibinfo {author} {\bibfnamefont {T.}~\bibnamefont {Sagawa}},\
  }{\bibfield  {journal} {\bibinfo
  {journal} {Phys. Rev. Lett.}\ }\textbf {\bibinfo {volume} {111}},\ \bibinfo
  {pages} {180603} (\bibinfo {year} {2013})}\BibitemShut {NoStop}%
\bibitem [{\citenamefont {Horowitz}\ and\ \citenamefont
  {Esposito}(2014)}]{HorowitzEsposito}%
  \BibitemOpen
  \bibfield  {author} {\bibinfo {author} {\bibfnamefont {J.~M.}\ \bibnamefont
  {Horowitz}}\ and\ \bibinfo {author} {\bibfnamefont {M.}~\bibnamefont
  {Esposito}},\ } {\bibfield  {journal}
  {\bibinfo  {journal} {Phys. Rev. X}\ }\textbf {\bibinfo {volume} {4}},\
  \bibinfo {pages} {031015} (\bibinfo {year} {2014})}\BibitemShut {NoStop}%
\bibitem [{\citenamefont {Hartich}\ \emph {et~al.}(2014)\citenamefont
  {Hartich}, \citenamefont {Barato},\ and\ \citenamefont {Seifert}}]{Hartich}%
  \BibitemOpen
  \bibfield  {author} {\bibinfo {author} {\bibfnamefont {D.}~\bibnamefont
  {Hartich}}, \bibinfo {author} {\bibfnamefont {A.~C.}\ \bibnamefont {Barato}},
  \ and\ \bibinfo {author} {\bibfnamefont {U.}~\bibnamefont {Seifert}},\
  } {\bibfield  {journal} {\bibinfo  {journal}
  {J. Stat. Mech.: Theor. Exp.}\ }\textbf {\bibinfo {volume} {2014}},\ \bibinfo
  {pages} {P02016} (\bibinfo {year} {2014})}\BibitemShut {NoStop}%
\bibitem [{\citenamefont {Munakata}\ and\ \citenamefont
  {Rosinberg}(2014)}]{Munakata}%
  \BibitemOpen
  \bibfield  {author} {\bibinfo {author} {\bibfnamefont {T.}~\bibnamefont
  {Munakata}}\ and\ \bibinfo {author} {\bibfnamefont {M.~L.}\ \bibnamefont
  {Rosinberg}},\ } {\bibfield  {journal}
  {\bibinfo  {journal} {Phys. Rev. Lett.}\ }\textbf {\bibinfo {volume} {112}},\
  \bibinfo {pages} {180601} (\bibinfo {year} {2014})}\BibitemShut {NoStop}%
\bibitem [{\citenamefont {Horowitz}\ and\ \citenamefont
  {Sandberg}(2014)}]{HorowitzSandberg}%
  \BibitemOpen
  \bibfield  {author} {\bibinfo {author} {\bibfnamefont {J.~M.}\ \bibnamefont
  {Horowitz}}\ and\ \bibinfo {author} {\bibfnamefont {H.}~\bibnamefont
  {Sandberg}},\ } {\bibfield  {journal} {\bibinfo  {journal} {New. J. Phys.}\
  }\textbf {\bibinfo {volume} {16}},\ \bibinfo {pages} {125007} (\bibinfo
  {year} {2014})}\BibitemShut {NoStop}%
\bibitem [{\citenamefont {Shiraishi}\ and\ \citenamefont
  {Sagawa}(2015)}]{ShiraishiSagawa}%
  \BibitemOpen
  \bibfield  {author} {\bibinfo {author} {\bibfnamefont {N.}~\bibnamefont
  {Shiraishi}}\ and\ \bibinfo {author} {\bibfnamefont {T.}~\bibnamefont
  {Sagawa}},\ } 
  {\bibfield  {journal} {\bibinfo  {journal} {Phys. Rev. E}\ }\textbf {\bibinfo
  {volume} {91}},\ \bibinfo {pages} {012130} (\bibinfo {year}
  {2015})}\BibitemShut {NoStop}%
\bibitem [{\citenamefont {Shiraishi}\ \emph {et~al.}(2015)\citenamefont
  {Shiraishi}, \citenamefont {Ito}, \citenamefont {Kawaguchi},\ and\
  \citenamefont {Sagawa}}]{ShiraishiItoKawaguchiSagawa}%
  \BibitemOpen
  \bibfield  {author} {\bibinfo {author} {\bibfnamefont {N.}~\bibnamefont
  {Shiraishi}}, \bibinfo {author} {\bibfnamefont {S.}~\bibnamefont {Ito}},
  \bibinfo {author} {\bibfnamefont {K.}~\bibnamefont {Kawaguchi}}, \ and\
  \bibinfo {author} {\bibfnamefont {T.}~\bibnamefont {Sagawa}},\ } {\bibfield  {journal}
  {\bibinfo  {journal} {New. J. Phys.}\ }\textbf {\bibinfo {volume} {17}},\
  \bibinfo {pages} {045012} (\bibinfo {year} {2015})}\BibitemShut {NoStop}%
\bibitem [{\citenamefont {Shiraishi}\ \emph {et~al.}(2016)\citenamefont
  {Shiraishi}, \citenamefont {Matsumoto},\ and\ \citenamefont
  {Sagawa}}]{ShiraishiMatsumotoSagawa}%
  \BibitemOpen
  \bibfield  {author} {\bibinfo {author} {\bibfnamefont {N.}~\bibnamefont
  {Shiraishi}}, \bibinfo {author} {\bibfnamefont {T.}~\bibnamefont
  {Matsumoto}}, \ and\ \bibinfo {author} {\bibfnamefont {T.}~\bibnamefont
  {Sagawa}},\ } {\bibfield  {journal} {\bibinfo  {journal} {New. J. Phys.}\ }\textbf
  {\bibinfo {volume} {18}},\ \bibinfo {pages} {013044} (\bibinfo {year}
  {2016})}\BibitemShut {NoStop}%
\bibitem [{\citenamefont {Koski}\ \emph {et~al.}(2015)\citenamefont {Koski},
  \citenamefont {Kutvonen}, \citenamefont {Khaymovich}, \citenamefont
  {Ala-N.},\ and\ \citenamefont {Pekola}}]{Koski2}%
  \BibitemOpen
  \bibfield  {author} {\bibinfo {author} {\bibfnamefont {J.~V.}\ \bibnamefont
  {Koski}}, \bibinfo {author} {\bibfnamefont {A.}~\bibnamefont {Kutvonen}},
  \bibinfo {author} {\bibfnamefont {I.~M.}\ \bibnamefont {Khaymovich}},
  \bibinfo {author} {\bibfnamefont {T.}~\bibnamefont {Ala-Nissila}}, \ and\ \bibinfo
  {author} {\bibfnamefont {J.~P.}\ \bibnamefont {Pekola}},\ }  {\bibfield  {journal}
  {\bibinfo  {journal} {Phys. Rev. Lett.}\ }\textbf {\bibinfo {volume} {115}},\
  \bibinfo {pages} {260602} (\bibinfo {year} {2015})}\BibitemShut {NoStop}%
\bibitem [{\citenamefont {Hartich}\ \emph {et~al.}(2016)\citenamefont
  {Hartich}, \citenamefont {Barato},\ and\ \citenamefont
  {Seifert}}]{HartichSeifertSensor}%
  \BibitemOpen
  \bibfield  {author} {\bibinfo {author} {\bibfnamefont {D.}~\bibnamefont
  {Hartich}}, \bibinfo {author} {\bibfnamefont {A.~C.}\ \bibnamefont {Barato}},
  \ and\ \bibinfo {author} {\bibfnamefont {U.}~\bibnamefont {Seifert}},\
  } {\bibfield  {journal} {\bibinfo  {journal} {Phys. Rev. E}\
  }\textbf {\bibinfo {volume} {93}},\ \bibinfo {pages} {022116} (\bibinfo
  {year} {2016})}\BibitemShut {NoStop}%
\bibitem [{\citenamefont {Barato}\ \emph {et~al.}(2013)\citenamefont {Barato},
  \citenamefont {Hartich},\ and\ \citenamefont {Seifert}}]{BaratoSeifert1}%
  \BibitemOpen
  \bibfield  {author} {\bibinfo {author} {\bibfnamefont {A.~C.}\ \bibnamefont
  {Barato}}, \bibinfo {author} {\bibfnamefont {D.}~\bibnamefont {Hartich}}, \
  and\ \bibinfo {author} {\bibfnamefont {U.}~\bibnamefont {Seifert}},\
  } {\bibfield  {journal} {\bibinfo  {journal} {Phys. Rev. E}\
  }\textbf {\bibinfo {volume} {87}},\ \bibinfo {pages} {042104} (\bibinfo
  {year} {2013})}\BibitemShut {NoStop}%
\bibitem [{\citenamefont {Barato}\ and\ \citenamefont
  {Seifert}(2013)}]{BaratoSeifert2}%
  \BibitemOpen
  \bibfield  {author} {\bibinfo {author} {\bibfnamefont {A.~Cardoso}\
  \bibnamefont {Barato}}\ and\ \bibinfo {author} {\bibfnamefont
  {U.}~\bibnamefont {Seifert}},\ } 
  {\bibfield  {journal} {\bibinfo  {journal} {Europhys. Lett.}\ }\textbf
  {\bibinfo {volume} {101}},\ \bibinfo {pages} {60001} (\bibinfo {year}
  {2013})}\BibitemShut {NoStop}%
\bibitem [{\citenamefont {Barato}\ and\ \citenamefont
  {Seifert}(2014)}]{BaratoSeifert3}%
  \BibitemOpen
  \bibfield  {author} {\bibinfo {author} {\bibfnamefont {A.~C.}\ \bibnamefont
  {Barato}}\ and\ \bibinfo {author} {\bibfnamefont {U.}~\bibnamefont
  {Seifert}},\ }  {\bibfield
  {journal} {\bibinfo  {journal} {Phys. Rev. E}\ }\textbf {\bibinfo {volume}
  {90}},\ \bibinfo {pages} {042150} (\bibinfo {year} {2014})}\BibitemShut
  {NoStop}%
\bibitem [{\citenamefont {Bauer}\ \emph {et~al.}(2012)\citenamefont {Bauer},
  \citenamefont {Abreu},\ and\ \citenamefont {Seifert}}]{BauerSeifert}%
  \BibitemOpen
  \bibfield  {author} {\bibinfo {author} {\bibfnamefont {M.}~\bibnamefont
  {Bauer}}, \bibinfo {author} {\bibfnamefont {D.}~\bibnamefont {Abreu}}, \ and\
  \bibinfo {author} {\bibfnamefont {U.}~\bibnamefont {Seifert}},\ }  {\bibfield  {journal} {\bibinfo  {journal} {J.
  Phys. A: Math. Theor.}\ }\textbf {\bibinfo {volume} {45}},\ \bibinfo {pages}
  {162001} (\bibinfo {year} {2012})}\BibitemShut {NoStop}%
\bibitem [{\citenamefont {Sandberg}\ \emph {et~al.}(2014)\citenamefont
  {Sandberg}, \citenamefont {Delvenne}, \citenamefont {Newton},\ and\
  \citenamefont {Mitter}}]{SandbergMitter}%
  \BibitemOpen
  \bibfield  {author} {\bibinfo {author} {\bibfnamefont {H.}~\bibnamefont
  {Sandberg}}, \bibinfo {author} {\bibfnamefont {J.~C.}\ \bibnamefont
  {Delvenne}}, \bibinfo {author} {\bibfnamefont {N.~J.}\ \bibnamefont
  {Newton}}, \ and\ \bibinfo {author} {\bibfnamefont {S.~K.}\ \bibnamefont
  {Mitter}},\ }\bibfield  {title}   {\bibfield  {journal} {\bibinfo  {journal} {Phys. Rev. E}\
  }\textbf {\bibinfo {volume} {90}},\ \bibinfo {pages} {042119} (\bibinfo
  {year} {2014})}\BibitemShut {NoStop}%
\bibitem [{\citenamefont {Jarillo}\ \emph {et~al.}(2016)\citenamefont
  {Jarillo}, \citenamefont {Tangarife},\ and\ \citenamefont
  {Cao}}]{JarilloCao}%
  \BibitemOpen
  \bibfield  {author} {\bibinfo {author} {\bibfnamefont {J.}~\bibnamefont
  {Jarillo}}, \bibinfo {author} {\bibfnamefont {T.}~\bibnamefont {Tangarife}},
  \ and\ \bibinfo {author} {\bibfnamefont {F.~J.}\ \bibnamefont {Cao}},\
  } {\bibfield  {journal}
  {\bibinfo  {journal} {Phys. Rev. E}\ }\textbf {\bibinfo {volume} {93}},\
  \bibinfo {pages} {012142} (\bibinfo {year} {2016})}\BibitemShut {NoStop}%
\bibitem [{\citenamefont {Demirel}\ and\ \citenamefont
  {Sandler}(2002)}]{Demirel}%
  \BibitemOpen
  \bibfield  {author} {\bibinfo {author} {\bibfnamefont {Y.}~\bibnamefont
  {Demirel}}\ and\ \bibinfo {author} {\bibfnamefont {S.~I.}\ \bibnamefont
  {Sandler}},\ } {\bibfield  {journal} {\bibinfo
  {journal} {Biophysical Chemistry}\ }\textbf {\bibinfo {volume} {97}},\
  \bibinfo {pages} {87} (\bibinfo {year} {2002})}\BibitemShut {NoStop}%
\bibitem [{\citenamefont {Sekimoto}(2005)}]{SekimotoSE}%
  \BibitemOpen
  \bibfield  {author} {\bibinfo {author} {\bibfnamefont {K.}~\bibnamefont
  {Sekimoto}},\ }
  {\bibfield  {journal} {\bibinfo  {journal} {Physica D}\ }\textbf {\bibinfo
  {volume} {205}},\ \bibinfo {pages} {242} (\bibinfo {year}
  {2005})}\BibitemShut {NoStop}%
\bibitem [{\citenamefont {Schaller}\ \emph {et~al.}(2010)\citenamefont
  {Schaller}, \citenamefont {Kie{\ss}lich},\ and\ \citenamefont
  {Brandes}}]{Schaller}%
  \BibitemOpen
  \bibfield  {author} {\bibinfo {author} {\bibfnamefont {G.}~\bibnamefont
  {Schaller}}, \bibinfo {author} {\bibfnamefont {G}~\bibnamefont
  {Kie{\ss}lich}}, \ and\ \bibinfo {author} {\bibfnamefont {T.}~\bibnamefont
  {Brandes}},\ }  {\bibfield  {journal} {\bibinfo  {journal} {Phys.
  Rev. B}\ }\textbf {\bibinfo {volume} {82}},\ \bibinfo {pages} {041303(R)}
  (\bibinfo {year} {2010})}\BibitemShut {NoStop}%
\bibitem [{\citenamefont {Thulasiraman}\ and\ \citenamefont
  {Swamy}(2011)}]{Graph}%
  \BibitemOpen
  \bibfield  {author} {\bibinfo {author} {\bibfnamefont {K.}~\bibnamefont
  {Thulasiraman}}\ and\ \bibinfo {author} {\bibfnamefont {M.~N.~S.}\
  \bibnamefont {Swamy}},\ }\href@noop {} {\emph {\bibinfo {title} {Graphs:
 Theory and Algorithms}}}\ (\bibinfo  {publisher} {John Wiley \& Sonsn, New York},\
  \bibinfo {year} {2011})\BibitemShut {NoStop}%
  
\end{thebibliography}
\end{document}